\documentclass[aps,prd,tightenlines,preprintnumbers,superscriptaddress,twocolumn,notitlepage,nofootinbib,eqsecnum,floatfix]{revtex4-1}
\usepackage[pdftex]{color}
\usepackage{hyperref}
\hypersetup{
    colorlinks=true,       
    linkcolor=blue,          
    citecolor=blue,        
    filecolor=blue,      
    urlcolor=blue           
}
\usepackage{amsmath}
\usepackage[utf8]{inputenc}
\usepackage{cases}
\usepackage{graphicx}
\usepackage{stackrel}
\usepackage{bm}

\usepackage{dcolumn}
  \newcolumntype{d}[1]{D{.}{.}{#1}}
  \newcolumntype{Y}[1]{D{.}{}{#1}}
  \newcolumntype{M}[1]{D{-}{-}{#1}}

\definecolor{red}{rgb}{0.8,0.0,0.0}
\definecolor{green}{rgb}{0.0,0.6,0.0}
\definecolor{darkblue}{rgb}{0.0,0.1,0.7}
\definecolor{brown}{rgb}{0.6,0.1,0.0}
\definecolor{grey}{rgb}{0.6,0.6,0.6}
\definecolor{darkgreen}{rgb}{0.0, 0.545098, 0.0}
\definecolor{applegreen}{rgb}{0.55, 0.71, 0.0}
\definecolor{purple}{rgb}{0.5,0.0,0.5}
\definecolor{babypink} {rgb}{0.64, 0.44, 0.44}
\definecolor{orange}{rgb}{1.0,0.5,0.0}

\newcommand{\amuL}{$a_\mu^{ll}(\mathrm{conn.})$}
\newcommand{\amuConn}{$a_\mu^{ud}({\rm conn.})$}
\newcommand{\PiOneL}{{$\Pi_1^{ll}({\rm conn.})$}}
\newcommand{\PiTwoL}{{$\Pi_2^{ll}({\rm conn.})$}}
\newcommand{\amu}{$a_\mu^{\rm HVP,LO}$}
\newcommand{\amuLval}{637.8(8.8)}
\newcommand{\amuTOTval}{699(15)_{u,d}(1)_{s,c,b}}
\newcommand{\amuDIFF}{1.3}
\newcommand{\PiOneLmath}{0.0932(14)}
\newcommand{\PiOneLval}{$\PiOneLmath\, {\rm GeV}^2$}
\newcommand{\PiTwoLmath}{-0.2089(64)}
\newcommand{\PiTwoLval}{$\PiTwoLmath\, {\rm GeV}^4$}
\newcommand{\amumath}{a_\mu^{\rm HVP,LO}}
\newcommand{\Pade}{Pad{\'e}}

\newcommand{\pipi}{$\pi\pi$}
\newcommand{\Pihat}{\widehat\Pi}
\newcommand{\alphaEM}{{\alpha}}
\newcommand{\gmtwo}{$g_\mu-2$}

\newcommand{\order}{\ensuremath{\text{O}}} 
\newcommand{\be}{\begin{equation}}
\newcommand{\ee}{\end{equation}}
\newcommand{\bea}{\begin{eqnarray}}
\newcommand{\eea}{\end{eqnarray}}

\newcommand{\ltapprox}{\lesssim} 
\newcommand{\gtapprox}{\gtrsim}  

\begin{document}
\preprint{FERMILAB-PUB-19-064-T}

\widetext

\title{Hadronic-vacuum-polarization contribution to the muon's anomalous magnetic moment from four-flavor lattice QCD }

\vskip 0.25cm

\author{C.~T.~H.~Davies}\email{christine.davies@glasgow.ac.uk}
\affiliation{SUPA, School of Physics and Astronomy, University of Glasgow, Glasgow, G12 8QQ, UK}

\author{C.~DeTar}
\affiliation{Department of Physics and Astronomy, University of Utah, \\ Salt Lake City, Utah, 84112, USA}

\author{A.~X.~El-Khadra}
\affiliation{Department of Physics, University of Illinois, Urbana, Illinois, 61801, USA}
\affiliation{Fermi National Accelerator Laboratory, Batavia, Illinois, 60510, USA}

\author{E.~G\'amiz}
\affiliation{CAFPE and Departamento de F\'{\i}sica Te\'orica y del Cosmos, Universidad de Granada,18071, Granada, Spain}

\author{Steven~Gottlieb}
\affiliation{Department of Physics, Indiana University, Bloomington, Indiana, 47405, USA}

\author{D.~Hatton}
\affiliation{SUPA, School of Physics and Astronomy, University of Glasgow, Glasgow, G12 8QQ, UK}

\author{A.~S.~Kronfeld}
\affiliation{Fermi National Accelerator Laboratory, Batavia, Illinois, 60510, USA}
\affiliation{Institute for Advanced Study, Technische Universit\"at M\"unchen, 85748 Garching, Germany}

\author{J.~Laiho}
\affiliation{Department of Physics, Syracuse University, Syracuse, New York, 13244, USA}

\author{G.~P.~Lepage}
\affiliation{Laboratory for Elementary-Particle Physics, Cornell University, Ithaca, New York 14853, USA}

\author{Yuzhi~Liu}
\affiliation{Department of Physics, Indiana University, Bloomington, Indiana, 47405, USA}

\author{P.~B.~Mackenzie}
\affiliation{Fermi National Accelerator Laboratory, Batavia, Illinois, 60510, USA}

\author{C.~McNeile}
\affiliation{Centre for Mathematical Sciences, {University of Plymouth}, Plymouth, PL4 8AA, UK}

\author{E.~T.~Neil}
\affiliation{Department of Physics, University of Colorado, Boulder, Colorado 80309, USA}

\author{T.~Primer}
\affiliation{Department of Physics, University of Arizona, Tucson, Arizona, 85721, USA}

\author{J.~N.~Simone}
\affiliation{Fermi National Accelerator Laboratory, Batavia, Illinois, 60510, USA}

\author{D.~Toussaint}
\affiliation{Department of Physics, University of Arizona, Tucson, Arizona, 85721, USA}

\author{R.~S.~\surname{Van de Water}}\email{ruthv@fnal.gov}
\affiliation{Fermi National Accelerator Laboratory, Batavia, Illinois, 60510, USA}

\author{A.~Vaquero}
\affiliation{Department of Physics and Astronomy, University of Utah, \\ Salt Lake City, Utah, 84112, USA}

\collaboration{Fermilab Lattice, HPQCD, and MILC Collaborations}
\noaffiliation

\date{February 18, 2020}

\begin{abstract}
We calculate the contribution to the muon anomalous magnetic moment hadronic vacuum polarization from {the} connected diagrams {of} up and down quarks{, omitting electromagnetism}.  We employ QCD gauge-field configurations with dynamical $u$, $d$, $s$, and $c$ quarks and the physical pion mass, and analyze five ensembles with lattice spacings ranging from $a \approx 0.06$ to~0.15~fm.  The up- and down-quark masses in our simulations have equal masses $m_l$.  We obtain{, in this} world where all pions have the mass of the $\pi^0${,} $10^{10}  a_\mu^{ll}({\rm conn.})  = \amuLval$, in agreement with independent lattice-QCD calculations.
We {then} combine this value with published lattice-QCD results for the connected contributions from strange, charm, and bottom quarks, and an estimate of the uncertainty due to the fact that our calculation does not include strong-isospin breaking, electromagnetism, or contributions from quark-disconnected diagrams.  Our final result for the total $\mathcal{O}(\alphaEM^2)$ hadronic vacuum polarization to the muon{'s anomalous magnetic moment} is~$10^{10} \amumath = \amuTOTval$, where the errors are from the light-quark and heavy-quark contributions, respectively.   Our result agrees with both {\it ab-initio} lattice-QCD calculations and phenomenological determinations from experimental $e^+e^-$-scattering data. {It} is $\amuDIFF\sigma$ below the ``no new physics" value of the hadronic-vacuum-polarization contribution {inferred from} combining the BNL E821 measurement of $a_\mu$ with theoretical calculations of the other contributions.
\end{abstract}

\pacs{}
\maketitle

\section{Introduction}
\label{sec:intro}

In the absence of direct evidence for new particles or forces
that are not present in the Standard Model, it becomes
increasingly important to pursue experiments that may yield indirect evidence.
Very heavy particles with masses beyond the reach of the Large Hadron
Collider can have a tiny effect on low-energy observables through
their brief appearance and disappearance in a quantum energy
fluctuation of the vacuum that couples to the observable.
Lighter particles with such small couplings to Standard-Model matter
that they have escaped detection could behave in a similar way.
To pin down such effects requires both very precise experimental
measurements and very good control of the theoretical calculations of the corresponding observables within the Standard-Model framework.

The anomalous magnetic moment of the muon, $a_{\mu}$, is such an observable.
It is defined as $(g_{\mu}-2)/2$ where the gyromagnetic ratio, $g_{\mu}$, which
connects the muon's spin and magnetic moment, would have a value of 2
in a world with no quantum corrections.
Consequently, the value of $a_{\mu}$ is sensitive to all of the particles that can
appear virtually in a quantum-field-theory description of the muon/photon
magnetic interaction.  Given a careful enumeration of all of the Standard-Model
contributions  to $a_{\mu}$, we can identify any significant discrepancy with experiment
as evidence for new physics.

The muon's anomalous magnetic moment was measured to an accuracy of 0.54 ppm nearly
20~years ago~\cite{Bennett:2006fi} at Brookhaven and will be updated to a planned accuracy of
0.14 ppm by the E989 experiment~\cite{Grange:2015fou,Hong:2018kqx} now running at Fermilab and the E34 experiment~\cite{Sato:2017sdn} still under development at J-PARC.
This prospect has galvanized a great deal of theoretical and linked experimental
activity to improve the accuracy of the Standard-Model result for $a_{\mu}$.
Recent calculations~\cite{Jegerlehner:2017lbd, Davier:2017zfy, Keshavarzi:2018mgv}
give a Standard-Model result with an uncertainty at 0.3 ppm
and a tantalizing 3.5--4\,$\sigma$ discrepancy with existing experiment.
This theoretical precision is sufficient to achieve a greater than $5\sigma$ significance for the
discrepancy if the central value does not change with the upcoming experimental
results. It is nevertheless important to test the uncertainty in the
Standard-Model result using different approaches to make sure that it is robust.

\begin{figure}[tb]
\centering
\includegraphics[width=0.2\textwidth]{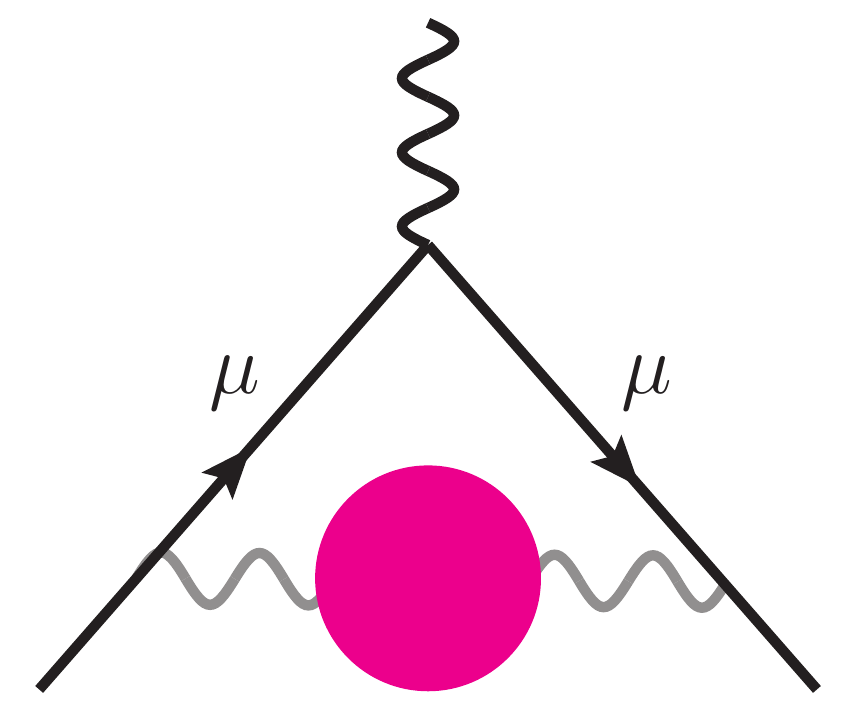}
\caption{
Leading hadronic contribution to the muon \gmtwo.    The shaded circle denotes {\it all} corrections to the internal photon propagator from the vacuum polarization of $u$, $d$, $s$, $c$, and $b$ quarks in the leading one-loop muon vertex diagram.  Diagrams in which the photon creates a quark-antiquark pair, which propagate while interacting via the strong and electromagnetic forces, and subsequently annihilate back into a photon, are called ``quark-connected" diagrams.  Those in which  the quark-antiquark pair annihilates into gluons are referred to as ``quark-disconnected" diagrams.}
\label{fig:HVP}
\end{figure}

The results of~Refs.~\cite{Jegerlehner:2017lbd, Davier:2017zfy, Keshavarzi:2018mgv}
use experimental input for the cross section for
$e^+e^-$ annihilation via a photon to hadrons as a function of center-of-mass energy
to determine an important hadronic contribution to $a_{\mu}$ known as
the leading-order hadronic vacuum polarization contribution, \amu.
This contribution, which appears at order $\alphaEM^2$, {where $\alphaEM$ is the fine structure constant,} is illustrated in Fig.~\ref{fig:HVP}. The uncertainty on its value is one of the two largest sources of error
in the Standard-Model result.
The leading-order hadronic vacuum polarization contribution can also be calculated from first principles using numerical lattice
QCD, and there has been a great deal of progress in the past few years on
improving lattice-QCD calculations of this quantity.\footnote{Another key uncertainty in the Standard-Model result comes from a higher-order hadronic piece known as the hadronic-light-by-light contribution. This is also being calculated in lattice QCD~\cite{Blum:2016lnc, Asmussen:2018ovy}.}
The aim of this effort is to reduce the uncertainty from lattice QCD {first} to a level
commensurate with that from using $\sigma(e^+e^- \rightarrow \text{hadrons})$, and then to the $\sim$0.2\% target precision of the Fermilab E989 and J-PARC experiments.
In the meantime, however, lattice-QCD calculations
already provide a strong test of those results from a
completely different method with very different systematic errors.

As illustrated in Fig.~\ref{fig:HVP}, \amu\ requires knowledge of the quark vacuum{-}polarization function that couples to a photon~\cite{Blum:2002ii, Lautrup:1971jf}.
In lattice QCD, individual diagrammatic contributions to the quark vacuum polarization can be considered separately via suitably constructed vector current-current correlation functions in Euclidean time. The vacuum polarization includes quark-line connected and disconnected diagrams, but the disconnected diagrams, where the quark loops are connected by intermediate gluons, contribute less than 2\% to \amu~\cite{Chakraborty:2015ugp,Blum:2015you,Borsanyi:2017zdw,Blum:2018mom,Shintani:2019wai,Gerardin:2019rua,Aubin:2019usy}. The quark-connected contribution can be further separated into contributions from the individual quark flavors, up, down, strange, charm, and bottom. Accurate lattice-QCD results for the separate $s$-, $c$- and (negligible) $b$-quark connected contributions to \amu\ were first obtained in Refs~\cite{Chakraborty:2014mwa, Donald:2012ga, Colquhoun:2014ica}. Subsequent lattice-QCD calculations~\cite{Blum:2016xpd, DellaMorte:2017dyu, Giusti:2017jof,Borsanyi:2017zdw,Blum:2018mom,Shintani:2019wai,Gerardin:2019rua} using different methods and quark formulations are in excellent agreement with these results.

The dominant quark-line connected contribution to \amu\ comes from the
light ($u/d$) quarks, however, and is the target of this work. Here lattice-QCD calculations carry
a number of additional technical challenges.
The vector current-current correlator falls more slowly with Euclidean time
at lighter quark masses, but at the same time the signal-to-noise degrades
more rapidly. This means that the light-quark connected contribution to \amu\ receives contributions from larger Euclidean times than those from heavy quarks and that the data at these times {are} noisier.  Hence controlling statistical errors is a challenge.
In addition large physical volumes are needed for the lattice-QCD calculation
to avoid systematic effects from squeezing light
states ({\it e.g.}, pions) into a small box.

The first lattice-QCD calculation of \amuConn, the light-quark connected contribution to \amu\, that included physical-mass $u/d$ quarks was presented in Ref.~\cite{Chakraborty:2016mwy}, followed by {several} other lattice-QCD results~\cite{DellaMorte:2017dyu, Borsanyi:2017zdw, Blum:2018mom,Shintani:2019wai,Gerardin:2019rua}.
{All of these results were obtained in the isospin-symmetric limit, but the calculations}
differ in the quark {formulation} used, the lattice
spacings and volumes available, and in the treatment of
statistical errors and finite-volume effects. The agreement between
different lattice-QCD calculations done independently will
in the end be an important test of the results.
Currently the lattice-QCD
results for \amuConn\ are spread over a range
of several percent, with uncertainties at the same level. These errors
are several times larger than those obtained
using the experimental information from cross sections for
$e^+e^- \rightarrow \text{hadrons}$. This means that
lattice-QCD calculations are not yet in a position to add
significant information to that available from
$e^+e^- \rightarrow \text{hadrons}$~\cite{Blum:2018mom}.
This first round of complete lattice-QCD calculations
has, however, crystallized the issues that must be addressed
to improve current results and ultimately reach the target experimental precision.

In this paper we present {a} calculation of the light-quark connected contribution to
\amu\ in the isospin-symmetric limit.
Like Ref.~\cite{Chakraborty:2016mwy}, our work uses the highly improved staggered quark (HISQ) action~\cite{Follana:2006rc} and MILC ensembles with four flavors of HISQ sea quarks~\cite{Bazavov:2012xda}. It also shares analysis strategies, a small set of common vector current-current correlator data, and {three} coauthors with Ref.~\cite{Chakraborty:2016mwy}.
Many improvements have been made, however, with respect to that work.
An important difference is that {\it all}  ensembles of gluon-field configurations used in our analysis include $u/d$ quarks of physical mass, eliminating the need for a chiral extrapolation, whereas in Ref.~\cite{Chakraborty:2016mwy} only two out of
ten ensembles were at physical $u/d$ quark mass.  In addition, we include ensembles at finer lattice spacings and one new ensemble that has approximately ten times the statistics of the others. The finer lattice spacings enable better control of the extrapolation to the continuum limit (zero lattice spacing), while the high-statistics ensemble allows
us to undertake a significant study of the signal-to-noise issue mentioned above.
This enables a better understanding of the impact of replacing
correlator data with parametrizations of that data at large
Euclidean times, and will be discussed further in Sec.~\ref{sec:Sims}.
There are also a number of differences in the analysis strategies employed in this work compared {with} Ref.~\cite{Chakraborty:2016mwy}, chiefly among them that the rescaling of the Taylor coefficients introduced in Ref.~\cite{Chakraborty:2016mwy} is not used here. A detailed discussion of our analysis, including the differences with Ref.~\cite{Chakraborty:2016mwy} is given in Secs.~\ref{sec:amuLat}, \ref{sec:FVDisc}, and \ref{sec:udConn}.

We {do} not present any new results for the contributions of strong-isospin-breaking
and QED effects to the leading-order hadronic vacuum polarization, nor
for quark-line disconnected contributions. Progress has been made on all these
small, but important, contributions recently~\cite{Chakraborty:2015ugp, Blum:2015you, Chakraborty:2017tqp, Borsanyi:2017zdw, Blum:2018mom, Giusti:2018vrc}. We summarize
the current situation for these pieces in Sec.~\ref{sec:Results},
to motivate the systematic uncertainty that we allow for not including them.

This paper is organized as follows. Section~\ref{sec:background} provides needed theoretical background to the calculation of the renormalized quark vacuum-polarization
function from lattice QCD that is the key ingredient in calculating the
leading-order hadronic vacuum polarization contribution to $a_{\mu}$.
Section~\ref{sec:Calculation} gives details of our numerical lattice-QCD calculation and the
methods we employ (along with data-driven tests of those methods)
to tackle the issues of the growth of statistical uncertainties
in the correlators and finite-volume effects.  Section~\ref{sec:Results} provides
our results for the light-quark connected contribution to \amu\ and for the slope and curvature of
the renormalized quark vacuum-polarization function, along with comprehensive error budgets for these quantities.
Finally, Sec.~\ref{sec:conclusions} gives our determination of the
total \amu\ from lattice QCD and compares it {with} other
lattice-QCD results. This section also discusses the prospects for further improvements from
lattice QCD that will allow significant input to be made to the
Standard-Model value for $a_{\mu}$ ahead of new experimental results.

\section{Background and methodology}
\label{sec:background}

The relation between the leading-order hadronic-vacuum-polarization contribution to the muon's anomalous magnetic moment and the renormalized quark vacuum-polarization function $\Pihat (Q^2) \equiv \Pi(Q^2) - \Pi(0)$, which is calculated here in lattice QCD, is given by~\cite{Blum:2002ii,Lautrup:1971jf}
\begin{equation}
\amumath = \left(\frac{\alphaEM}{\pi}\right)^2 \int_0^{\infty} dQ^2 K_E(Q^2) \Pihat(Q^2) \,,
\label{eq:amuHVP}
\end{equation}
where $Q$ denotes the Euclidean momentum carried by the virtual photons and $K_E(Q^2)$ is the standard kernel function introduced by Blum in Ref.~\cite{Blum:2002ii}.  The integrand peaks around $Q^2 \approx m^2_{\mu}/2$.

The light-quark connected contribution to the muon's anomalous magnetic moment, \amuConn, arises from diagrams in which the photon in Fig.~\ref{fig:HVP} produces light $u\bar u$ or $d\bar d$ pairs.
We therefore start our lattice-QCD calculation of \amuConn\ with the zero-momentum $u/d$-quark current-current correlation function in Euclidean space,
\be
G(t) = \frac{1}{27} \int d{\mathbf{x}} \left[4 \langle j^u_i(\mathbf{x},t)j^u_{i}(0,0) \rangle
+  \langle j^d_i(\mathbf{x},t)j^d_{i}(0,0) \rangle \right],
\ee
where the summed index $i$ runs over spatial components and $j_{i}^q = \bar{q} \gamma_i q$.  The factor of 4 in front of the first term arises from the ratio of the quarks' electric charges squared, $q_u^2 / q_d^2$.
Following Ref.~\cite{Chakraborty:2014mwa}, we first compute time moments of $G(t)$, which are proportional to the coefficients $\Pi_{j}$ in a Taylor expansion of $\Pihat(Q^2)$ around $Q^2=0$. We then obtain $\Pihat(Q^2)$ from $[n,n]$ and $[n,n-1]$ \Pade\ approximants with $n=3$.
{Because $\Pihat(Q^2)$ can be expressed in terms of a Stieltjes integral through a once-subtracted dispersion relation (see, {\it e.g.}, Ref.~\cite{Aubin:2012me}),} the true result for $\Pihat(Q^2)$ is guaranteed to lie between the $[n,n]$ and $[n,n-1]$ \Pade\ approximants~\cite{Baker:1969sw,Barnsley:1973zi}. We find that the systematic uncertainty on \amu\ from the use of \Pade\ approximants decreases with increasing $n$, and is negligible even compared with the target experimental uncertainty for $n \geq 3$.
Indeed, we have checked with our lattice correlation functions that the time-moment method with $n=3$ \Pade\ approximants as used in this work yields results  for \amuConn\ that are numerically equivalent (to two decimal places or better) to the method introduced by Bernecker and Meyer in Ref.~\cite{Bernecker:2011gh} based on the  time-momentum representation of the Euclidean vector-current correlator.

In the time-momentum representation, $\Pihat$ is obtained directly from $G(t)$ via the integral~\cite{Bernecker:2011gh}
\begin{eqnarray}
\Pihat(\omega^2) & = & \frac{4\pi^2}{\omega^2} \int_0^{\infty} dt\, G(t) \left[ \omega^2 t^2 - 4 \sin^2\left(\frac{\omega t}{2}\right) \right],\quad \label{eq:PiHatLat}
\end{eqnarray}
which is a simpler procedure than calculating the \Pade\ approximants from the time-moments of $G(t)$.
However, the time-moment method directly yields the Taylor coefficients, and hence allows us to correct them for finite-volume and lattice discretization effects using a chiral model of pions and $\rho$ mesons {\em before} constructing \amuConn. In practice, then, the uncorrected values of \amuConn\ reported in Sec.~\ref{sec:Calculation} use Eq.~(\ref{eq:PiHatLat}) above, while the Taylor coefficients and corrected values of \amuConn\ are obtained from the time-moment method with $n=3$ \Pade\ approximants.

The traditional and currently still most precise determinations of \amu\ use dispersive methods to obtain the vacuum-polarization function from experimental ``$R$-ratio" data~\cite{Kurz:2015fhj,Jegerlehner:2017lbd,Davier:2017zfy,Keshavarzi:2018mgv}
\begin{equation}
\Pihat(Q^2) = \frac{Q^2}{3} \int_0^\infty ds\, \frac{R_\gamma(s)}{s(s+Q^2)} \,, \label{eq:PihatToR}
\end{equation}
with
\begin{equation}
R_\gamma(s) \equiv \frac{\sigma(e^+ e^- \to \gamma^* \to {\rm hadrons})}{4\pi\alpha(s)^2 / (3s)} \,,
\label{eq:Rratio}
\end{equation}
where $s$ is the square of the center-of-mass energy.  With this approach, one integrates over all hadronic channels and it is not possible to cleanly identify which light-quark flavor was created
at the photon vertex.  Hence, one cannot separate their contributions to
the cross section.  One can, however, isolate heavy-quark contributions to the $e^+e^-$ cross section
(see, e.g., Ref.~\cite{Chetyrkin:2009fv}), enabling a clean comparison
between lattice QCD and phenomenology.
This is most clearly done at the level of the Taylor coefficients of the contribution to
$\Pihat(Q^2)$ for that quark flavor. The good agreement seen between lattice-QCD
$c$- and $b$-quark connected contributions to $\Pi_j$ and those from $\sigma(e^+e^- \rightarrow {\text{hadrons}})$~\cite{Donald:2012ga, Colquhoun:2014ica, Nakayama:2016atf} further substantiates the methods employed in the lattice-QCD calculations.
In Sec.~\ref{sec:FVDisc}, we compare our lattice-QCD calculations of the Taylor
coefficients $\Pi_{j}$ summed over all flavors with those from $R$-ratio data to check
our model for calculating corrections due to nonzero lattice spacing and finite spatial volume.

\section{Lattice-QCD calculation}
\label{sec:Calculation}

We now present our lattice-QCD calculation.  First, in Sec.~\ref{sec:Sims}, we describe the numerical simulations.  We present the lattice quark and gluon actions employed and the parameters of the QCD gauge-field configurations and correlation functions.  Next, in Sec.~\ref{sec:amuLat}, we extract \amuConn\ in the isospin-symmetric limit
on each ensemble from the vector-current correlation functions. We describe our approach for dealing with the substantial statistical noise in our two-point correlators at large times.  Because we adapt many of the strategies of Ref.~\cite{Chakraborty:2016mwy} in our analysis, we highlight key differences and improvements with respect to that work.  Last, in Sec.~\ref{sec:FVDisc}, we correct the results for
the isospin-symmetric \amuConn\ on each ensemble for finite-volume and taste-breaking discretization errors, and subsequently extrapolate these corrected values to zero lattice spacing.

\subsection{Numerical simulations}
\label{sec:Sims}

We perform our calculation on QCD gauge-field configurations generated by the MILC Collaboration with four flavors of HISQ quarks~\cite{Follana:2006rc,Bazavov:2012xda}.
These configurations are isospin-symmetric, {\it i.e.}, the up and down sea-quark masses are equal with a mass $m_l  = (m_u + m_d)/2$.  We employ five ensembles with lattice spacings spanning $a\approx 0.15$--0.06 fm and physical-mass light, strange, and charm sea quarks.
The spatial volumes satisfy $M_\pi L \gtapprox 3.3$ with $M_\pi$ the taste-Goldstone pion mass, while the temporal extents range (from coarsest to finest lattice spacing) between $ 7.2 \gtapprox T \gtapprox 10.2$~fm.
Table~\ref{tab:ensembles} summarizes key parameters of the configurations.  Because our simulation light-quark masses are degenerate, throughout this work we use \amuL\ to denote the quark-connected contribution from two light flavors in the isospin-symmetric limit.  We reserve the notation \amuConn\ for nature's value.

\begin{table*}[tbh]
    \caption{Parameters of the QCD gauge-field ensembles.  The first column shows the approximate lattice spacing, while the second lists the bare lattice up, down, strange, and charm sea-quark masses.  The third column gives the ratio of the lattice spacing to the gradient-flow scale $w_0$~\cite{Borsanyi:2012zs}; to convert quantities in lattice-spacing units to GeV, we use $w_0=0.1715(9)$~fm~\cite{Dowdall:2013rya}.   The fourth column gives the nonperturbatively determined vector current renormalization factor obtained (for $s$ quarks) in Ref.~\cite{Chakraborty:2017hry}.  The fifth column lists the taste-Goldstone sea-pion masses; these were obtained from fits of pseudoscalar-current two-point correlators as in Ref.~\cite{Bazavov:2012xda}.  The sixth column shows the lowest-lying noninteracting two-pion energy level that couples to our vector current on each ensemble.  The seventh column gives the lattice volumes. The final two columns give the number of configurations analyzed and the number of random-wall time sources used per configuration, where ``TSM" indicates that we used the truncated solver method on this ensemble. \vspace{1mm}}
    \label{tab:ensembles}
\begin{ruledtabular}
\begin{tabular}{lcccccccc}
$\approx a$ (fm) & $am_l^{\rm sea}/am_s^{\rm sea}/am_c^{\rm sea}$ & $w_0/a$ & $Z_{V,\bar{s}s}$ & $M_{\pi_5}$ (MeV) & {$E_{2\pi,{\rm min}}$ }(MeV) & $(L/a)^3 \times (T/a)$ & $N_{\rm conf.}$  & $N_{\rm wall}$ \\
\hline
0.15 & 0.00235/0.0647/0.831 & 1.13670(50) & 0.9881(10) & 133.04(70) & {640.4(3.4)} & $32^3 \times 48$ & 997 & 16 \\
0.15 & 0.002426/0.0673/0.8447 & 1.13215(35) & 0.9881(10) & 134.73(71) & {639.7(3.4)} & $32^3 \times 48$ & 9362 & 48 (TSM) \\
0.12 & 0.00184/0.0507/0.628 & 1.41490(60) & 0.99220(40) & 132.73(70) & {540.8(3.3)} & $48^3 \times 64$ & 998 & 16 \\
0.09 & 0.00120/0.0363/0.432 & 1.95180(70) & 0.99400(50) & 128.34(68) & {524.3(2.8)} & $64^3 \times 96$ & 1557 & 16 (TSM) \\
0.06 & 0.0008/0.022/0.260 & 3.0170(23) & 0.9941(11) & 134.95(72) & {530.8(2.8)} & $96^3 \times 192$ & 1230 & 16 (TSM) \\
\end{tabular}

\end{ruledtabular}
\end{table*}

Two of the ensembles listed in Table~\ref{tab:ensembles} were also used in Ref.~\cite{Chakraborty:2016mwy}: the $a\approx 0.15$~fm ensemble with approximately 1000 configurations and the $a\approx 0.12$~fm ensemble.
Our analysis includes two new ensembles with $a\approx 0.09$~fm and $a\approx 0.06$~fm; the latter has a finer lattice spacing than those employed in Ref.~\cite{Chakraborty:2016mwy}, thereby providing better control over the continuum extrapolation.
In addition, a new ensemble is included with $a\approx 0.15$~fm and parameters identical to the older $a\approx 0.15$~fm physical-mass ensemble, except for having
better tuned quark masses. The new ensemble has 10,000 configurations, which is a factor
of ten better statistics.
On this ensemble, we can obtain \amuL\ to high precision directly from the lattice vector-current correlator as described in Sec.~\ref{sec:background}.
Thus, comparing this high-statistics ensemble and the older low-statistics one enables us to test our methods for extracting \amuL\ from noisy data.
Because we employ only physical-mass ensembles, a chiral extrapolation is not needed.

Following Ref.~\cite{Chakraborty:2016mwy}, on each ensemble we construct zero-momentum vector-current
correlators with the valence-quark mass equal to the light sea-quark mass and four combinations
of local and spatially smeared interpolating operators at the source and sink.
We use the taste-vector current that combines quark and antiquark propagators at a single lattice site.
The spatially smeared interpolating operators have the same taste because we employ a smearing function
that combines separations of an even number of lattice spacings. This function is given
in Eq.~(A1) of Ref.~\cite{Chakraborty:2016mwy}, where the smearing parameters are also listed
for lattice spacings $a\approx 0.15$--0.09~fm. For the $a \approx 0.06$~fm ensemble, we use a
smearing radius that is the same in physical units as the one employed at $a\approx 0.09$~fm, which
yields the smearing parameters $r_0 = 6.75$ and $n = 100$.
The correlators with smeared interpolating operators improve our identification of low-lying energy levels,
to be discussed in Sec.~\ref{sec:amuLat}.
We take the correlators on the low-statistics $a\approx 0.15$~fm and $a\approx 0.12$~fm ensembles directly from Ref.~\cite{Chakraborty:2016mwy}.  These correlators were computed with 16 equally spaced random-wall time sources and averaged to gain statistics.

On the three newer ensembles analyzed in this work, we employ in addition a cost-effective variance-reduction technique called the truncated solver method (TSM)~\cite{Blum:2012uh}.  With this approach, on each configuration we compute a large number of ``sloppy" correlators with a large relative error of $10^{-5}$ at a small cost, and a single ``fine" correlator with a small relative error of $10^{-8}$.  We correct the average of the sloppy results using the difference between the approximate and precise solutions on a single {source}.   In practice, we calculate sloppy propagators with all 48 time sources on the high-statistics $a\approx 0.15$~fm ensemble, and from 16 time sources on the $a\approx 0.09$ and $0.06$~fm ones.  Use of the TSM reduces our computational cost by more than a factor of~2.

\subsection{Extraction of muon anomaly}
\label{sec:amuLat}

\begin{figure}[tb]
\centering
\includegraphics[width=0.4\textwidth]{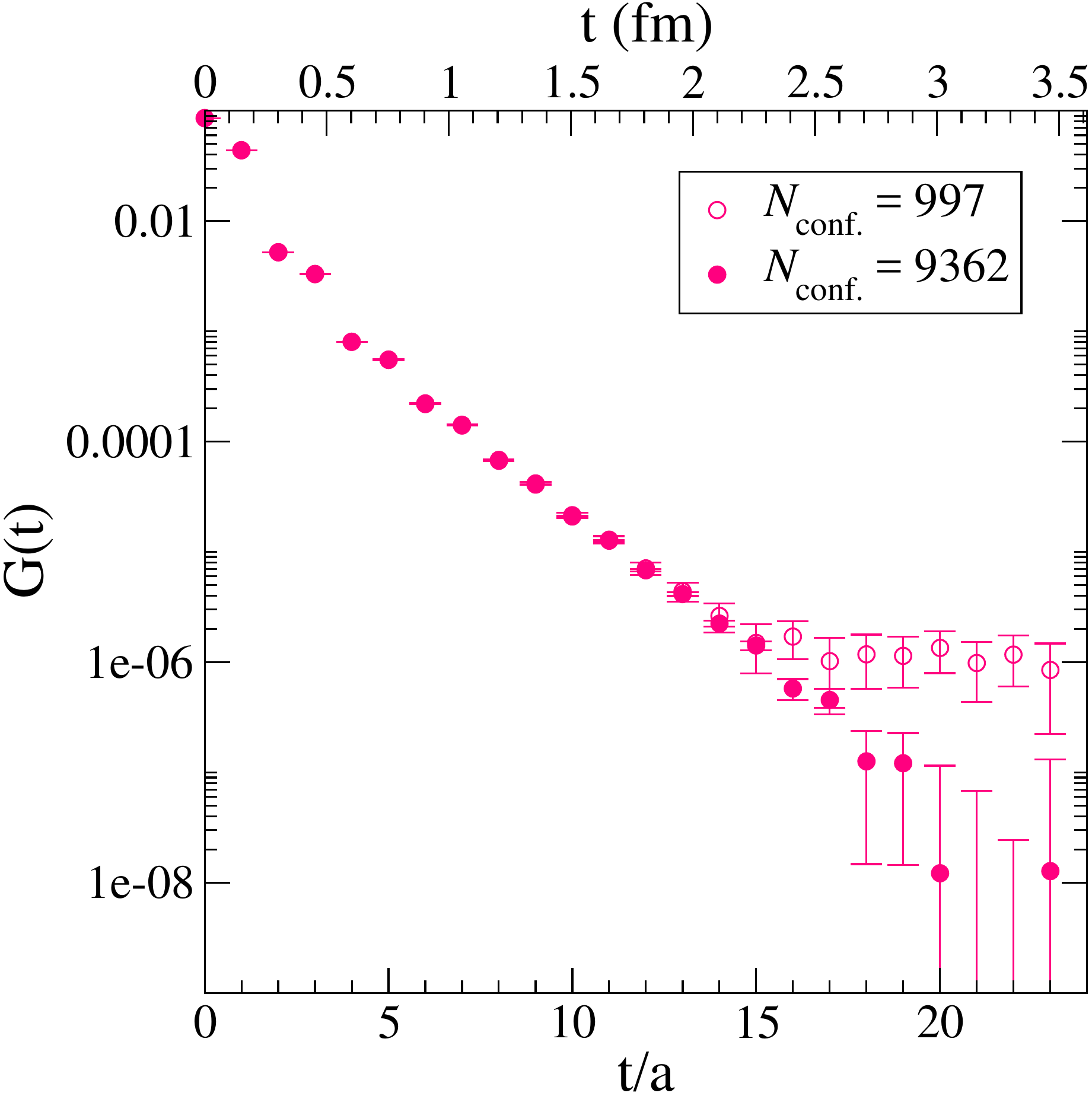}
\caption{ Local-local vector-current correlator on the two $a\approx 0.15$~fm ensembles with similar parameters but differing statistics.  Based on this plot, we choose $t_{\rm max}/a = 15$ and $t_{\rm max}/a = 24$ for the correlator fits on the low- and high-statistics ensembles, respectively.  Plots for other ensembles look similar to the low-statistics $a\approx 0.15$~fm data.}
\label{fig:vcoarse-Gt}
\end{figure}

A challenge common to all lattice-QCD calculations of \amuL\ is the large statistical noise in the vector-current correlator at the physical light-quark mass, in particular for distances above about 2--3~fm.
Figure~\ref{fig:vcoarse-Gt} shows the local-local vector-current correlator $G(t)$ on the two $a\approx 0.15$~fm ensembles.  We average the correlator values at times $t$ and $T-t$ to increase statistics, and thus show the correlator only up to the lattice temporal midpoint. The low- and high-statistics data agree for times below 2~fm.  Beyond this range, the data with low statistics become too noisy to yield a reliable estimate of the correlation function, and hence of the contribution to \amuL\ from large times.

Several strategies to address the noise problem have been used in the literature~\cite{DellaMorte:2017dyu,Borsanyi:2016lpl,Blum:2018mom}; here we follow the strategy of Ref.~\cite{Chakraborty:2016mwy}.  We first fit the $2\times 2$ matrix of correlators with combinations of local and smeared sources and sinks together using the parametrization in Eq.~(A2) of~\cite{Chakraborty:2016mwy}, constraining the energies and amplitudes with the Gaussian priors given in Eqs.~(A3) and~(A4) of~\cite{Chakraborty:2016mwy}. {In these fits, we minimize an augmented $\chi^2$ that includes contributions from both the data and the priors~\cite{Lepage:2001ym}.  Our} fit function is simply a sum of exponentials $\exp(-Et)$ such that the lowest-energy states are the only ones that survive to large time.  With staggered quarks, the two-point correlators receive contributions from both correct parity and opposite-parity states; the latter lead to contributions that oscillate with time as $(-1)^t$.  For every normal state in our fit ($N_{\rm states}$), we also include an opposite parity state.
We then replace the local-local correlator data for times {above a chosen time $t^*$} by the result of the multiexponential fit, and use this mixed data + fit correlator to calculate \amuL\, either via \Pade\ approximants or the time-momentum representation.  Our detailed fit choices, {e.g.} fit ranges and number of states included,  are given in Table~\ref{tab:2pt-fits}. They differ slightly from those of Ref.~\cite{Chakraborty:2016mwy}.

One must be careful with directly using the noisy large-time correlator data to calculate \amuL\,\!.
For all ensembles, we fix the maximum time ($t_{\rm max}$) included in the fit based on plots of the
local-local correlator (see Fig.~\ref{fig:vcoarse-Gt}), choosing $t_{\max}$
slightly below the time at which the $G(t)$ stops decaying exponentially.
Beyond this point, the data violate the model-independent upper bound pointed out in Ref.~\cite{Borsanyi:2016lpl} that $G(t)$ must fall off more rapidly than $\exp(-E_{\pi\pi}t)$, where $E_{\pi\pi}$ is the energy of two pions each with the smallest nonvanishing lattice momentum. The correlators stop decaying exponentially at around 2.3--2.6~fm on all ensembles with $\sim$1000 configurations.  In constrast, the correlator on the ensemble with almost 10,000 configurations displays an exponential (cosh) fall-off until the lattice midpoint.

\begin{figure}[tb]
\centering
\includegraphics[width=0.4\textwidth]{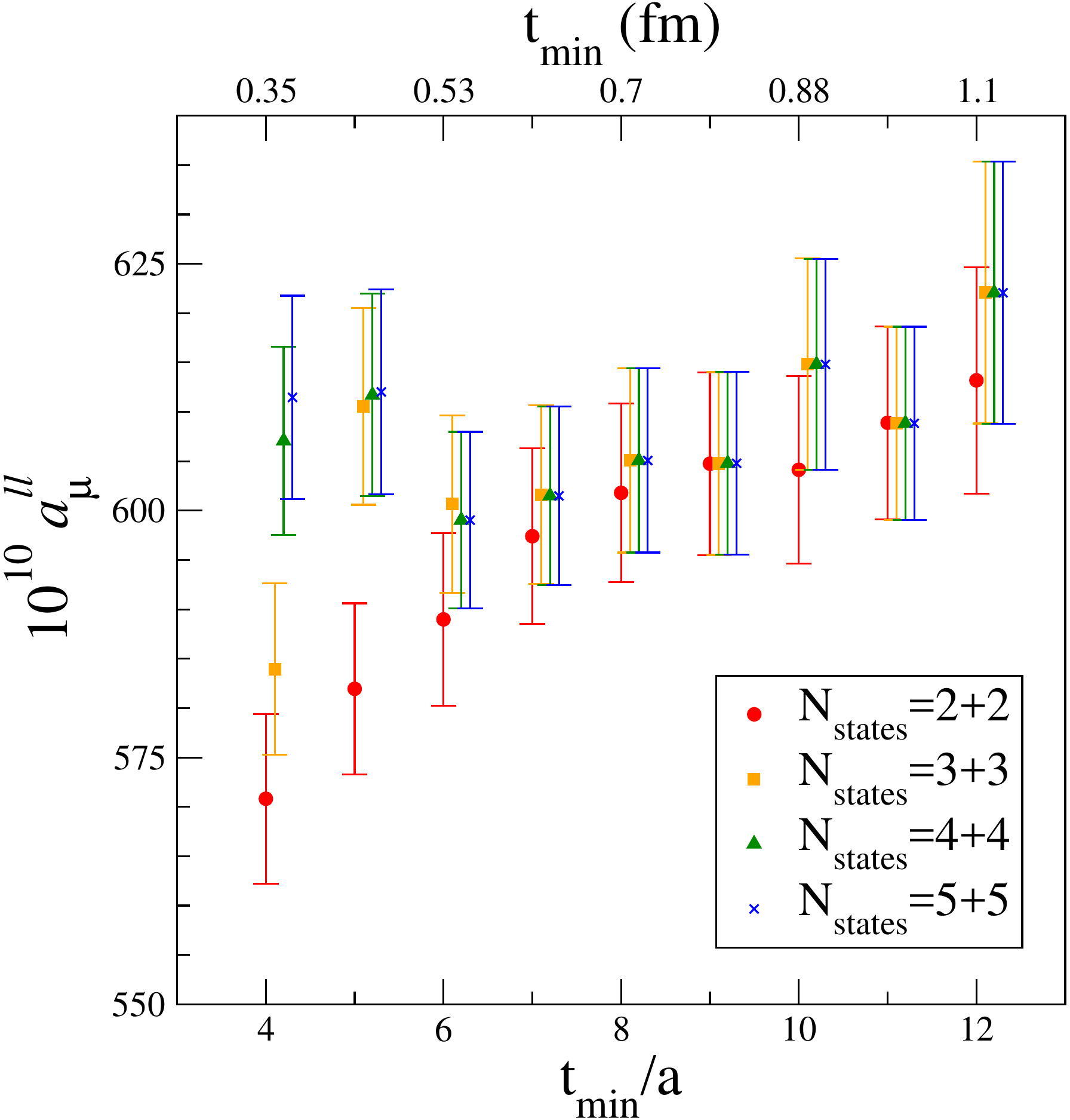}
\caption{ Stability of \amuL\ calculated from the mixed correlator $G_{\rm data}(t\leq 2.0$~fm) and $G_{\rm fit}(t > 2.0$~fm) on the $a\approx 0.09$~fm ensemble.  For each value of $t_{\rm min}$, the results for \amuL\ from fits with 2--5 pairs of oscillating and nonoscillating states are shown with a slight horizontal displacement for clarity; $t_{\rm max}/a=30$ for all fits.  For this ensemble, we select $t_{\rm min}/a = 8$ and three pairs of states. }
\label{fig:fine-amu-stab}
\end{figure}

After fixing $t_{\rm max}$, we then vary the minimum time in the fit range ($t_{\rm min}$) and the number of states in the fit function ($N_{\rm states}$) and look for good correlated fits with stable central values and errors.  Figure~\ref{fig:fine-amu-stab} plots \amuL\ versus $t_{\rm min}$ and $N_{\rm states}$ on the $a\approx 0.09$~fm ensemble.  The inclusion of more states in the fit improves fits with smaller minimum times, and the \amuL\ determinations are roughly independent of $t_{\rm min}$ and $N_{\rm states}$ for $t_{\rm min}/a \gtapprox 8$.  The stability plots for other ensembles are qualitatively similar.  Based on these plots, we choose $t_{\rm min} = 0.6$~fm on the $a \approx 0.15$~fm ensembles, and increase $t_{\rm min}$ smoothly with decreasing lattice spacing to $t_{\rm min} = 0.73$~fm on the $a \approx 0.06$ fm ensemble.

The true spectrum of the vector-current correlators is more complicated than the simple fit parametrization employed in our analysis, with many more levels than can be resolved within our finite statistics.  Although we cannot identify the asymptotic lowest \pipi\ energy level due to the large statistical noise in our data above around 2.5--3 fm, we can infer the presence of low-lying \pipi\ states from the fitted ground-state energies, which are below the $\rho$ pole on the finer ensembles.   Even with these caveats, however, our fits provide a sufficiently accurate extrapolation of $G(t)$ for the purposes of obtaining \amuL\,\!.  We have tested our noise-reduction strategy in several ways, and summarized the studies that provide the strongest substantiation of our approach below.

\begin{table*}[tb]
    \caption{Parameters of the vector-current correlator fits.   Ensembles are listed in the same order as in Table~\ref{tab:ensembles}.  The number of degree-of-freedom is $3\times$ the number of time slices in the fit range, rather than 4, because we average the local-source/smeared-sink and smeared-source/local-sink correlators (which should be equal in the limit of infinite statistics) before fitting. {The last column shows the standard frequentist $p$-values calculated from the $\chi^2$ contribution from the data only ($\chi^2_{\rm data}$), and with the degrees-of-freedom equal to the number of data points minus the number of fit parameters.} \vspace{1mm}}
    \label{tab:2pt-fits}
\begin{ruledtabular}
\begin{tabular}{lccd{4.8}c}
$\approx a$ (fm) & $[t_{\rm min},t_{\rm max}]/a$ & $N_{\rm states}$ & \multicolumn{1}{c}{$\chi^2_{\rm data}$/d.o.f.  [d.o.f.]} & $p$ \\
\hline
0.15 & [4,15] & 3+3 & 0.{90[18]} & 0.60\\
0.15 & [4,24] & 4+4 & 1.{22[39]} & 0.17\\
0.12 & [5,20] & 3+3 & 0.{75[30]} & 0.86\\
0.09 & [8,30] & 3+3 & 1.{42[51]} & 0.04\\
0.06 & [13,40] & 3+3 & 1.{16[66]} & 0.28\\
\end{tabular}
\end{ruledtabular}
\end{table*}

\begin{figure}[tb]
\centering
\includegraphics[width=0.4\textwidth]{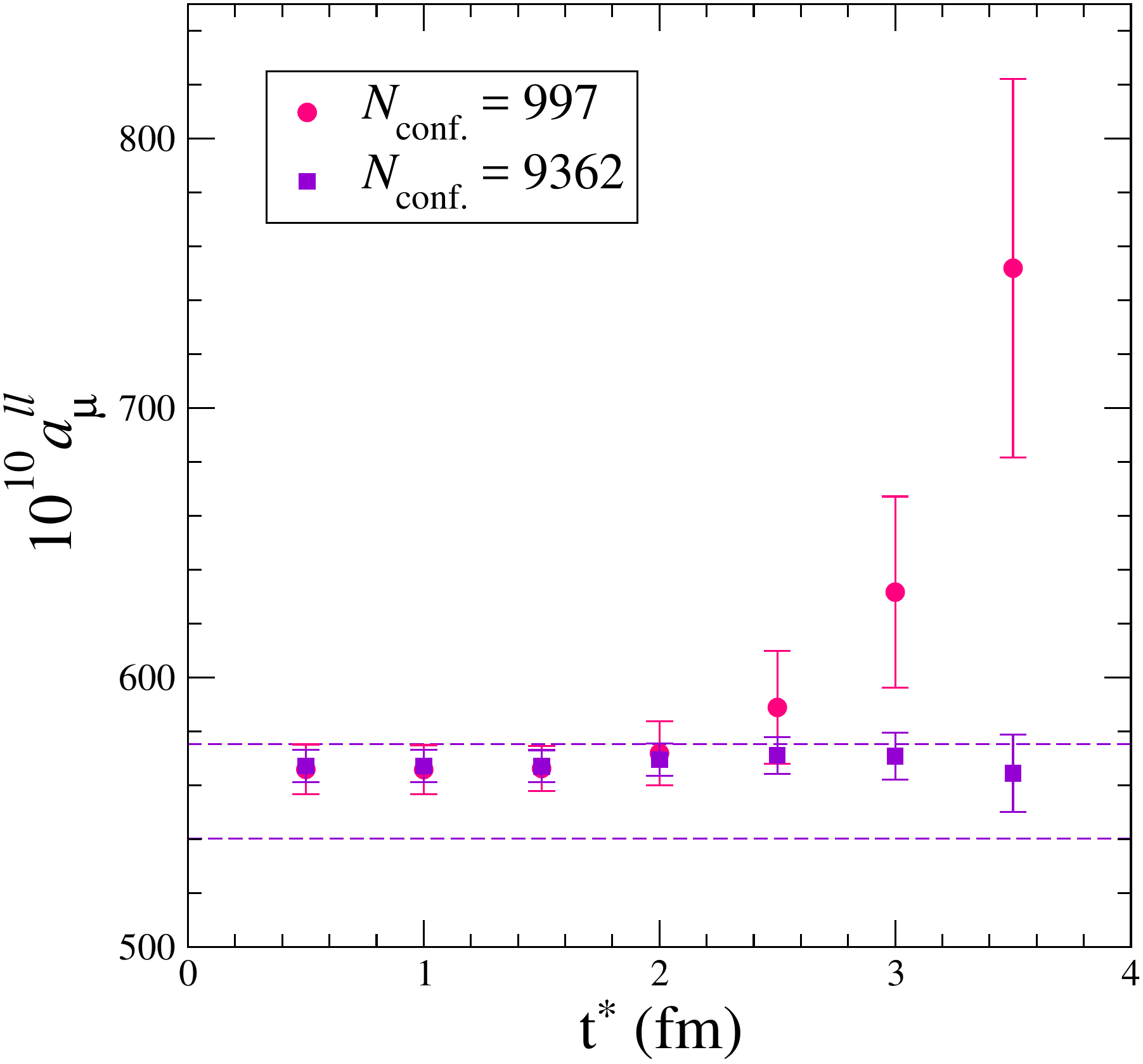}
\caption{\amuL\ versus {the transition time} $t^*$ {in the mixed data+fit correlator} on the two ensembles with $a \approx 0.15$~fm.  The dashed horizontal lines show the $1\sigma$ error band for the value of \amuL\ calculated entirely from data on the high-statistics ensemble.}
\label{fig:amu_vs_tstar}
\end{figure}

The $a\approx 0.15$~fm ensemble with $\sim$10,000 configurations enables a test of our use of correlator fits because on this ensemble we can obtain \amuL\ reliably from data alone.
Figure~\ref{fig:amu_vs_tstar}, left, shows the dependence of \amuL\ computed from the mixed correlator on $t^*$ in fm for the two $a\approx 0.15$~fm ensembles.  Also shown is the $1\sigma$ error band for the value of \amuL\ calculated entirely from data on the high-statistics ensemble.
We find that, for all times $t^* \ltapprox 2.5$~fm, the results for \amuL\ obtained from $G_{\rm data}(t\leq t^*)$ and $G_{\rm fit}(t>t^*)$ are consistent with the high-statistics data value.  Further, the results on the low- and high-statistics ensembles are consistent with each other.
This demonstrates that the fitted correlator yields an accurate value for \amuL\ provided  $t^* \ltapprox 2.5$~fm.

\begin{figure}[tb]
    \centering
    \includegraphics[width=0.44\textwidth]{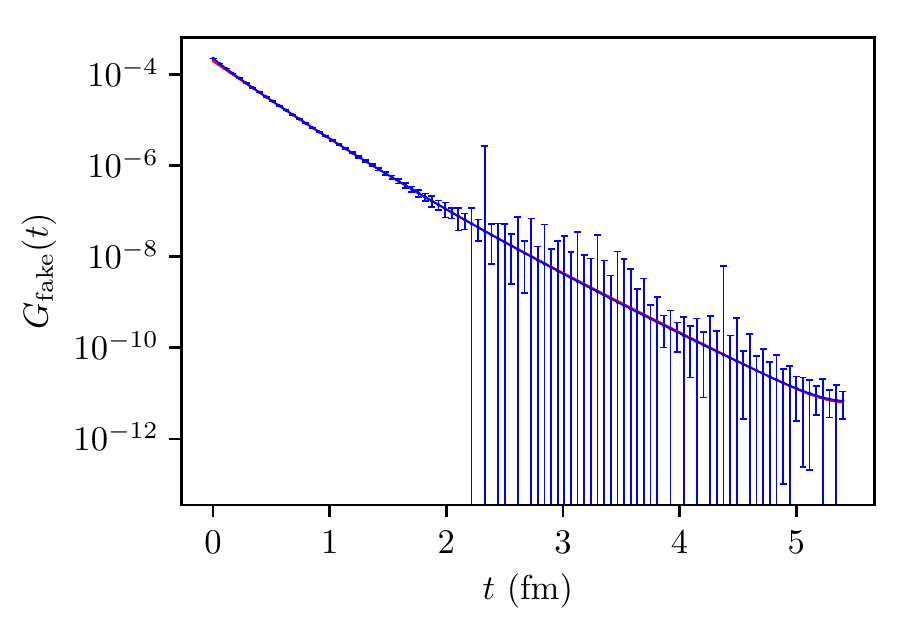} \vspace{-0.07in}
    \caption{{Time-}dependence of fake-data {correlator} $G_\mathrm{fake}(t)$ (blue) created from the chiral model used to calculate finite-volume corrections for $a=0.06$\,fm simulations {compared with the} result from a least-squares fit with two exponentials (red). {The agreement between the data and fit is so close that the blue curve obscures the red curve over most of the figure.}}
    \label{fig:model_vs_fit_G}
\end{figure}

The number of low-lying \pipi\ states in our vector-meson correlators increases rapidly as the lattice spacing, and consequently the taste splittings between sea-pion masses, decreases.
Thus, it is also important to test our use of correlator fits with data that have several states below the $\rho$.
To obtain a correlator similar to our $a\approx 0.06$~fm lattice data, but for which we know the spectrum exactly, we employ the chiral model in Appendix~B of Ref.~\cite{Chakraborty:2016mwy}.
We first calculate the finite-volume energy levels, including $\rho$-\pipi\ interactions, up to 2~GeV for our finest lattice spacing, $a\approx 0.06$~fm.  We then construct a fake correlator {$G_{\rm fake}(t)$} with central values computed from the approximately 30 model energies and amplitudes, and a covariance matrix obtained from the simulation correlator $G_{\rm data}(t)$.
We {then} fit $G_{\rm fake}(t)$ using the same fit range as in our analysis, and two or more states.  {Figure~\ref{fig:model_vs_fit_G} plots $G_{\rm fake}(t)$ along with the result of a two-exponential fit.}

\begin{figure}[tb]
\centering
\includegraphics[width=0.44\textwidth]{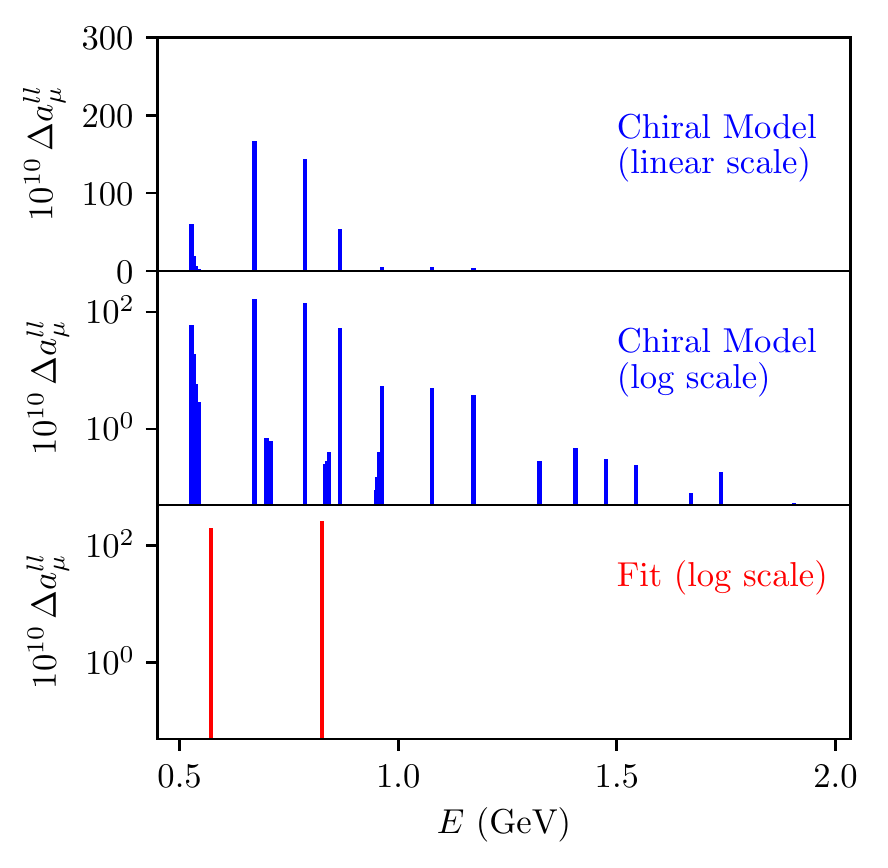} \vspace{-0.04in}
\caption{Top two panels: contributions to \amuL\ from all $\pi\pi$ states with energies~$E<2$\,GeV
in the chiral model used to calculate finite-volume corrections for the $a=0.06$\,fm simulations.
The top panel uses a linear $y$-scale; the second panel uses a $\log$~scale so
that a more complete set of energy levels can be displayed.
Bottom panel: contributions to \amuL\ from the states in a two-exponential fit to the fake data
created from the chiral model. Summing all contributions in each case gives results that agree to
within $0.9\times 10^{-10}$, which is roughly 1/10 the fit error.}
\label{fig:model_vs_fit_amu}
\end{figure}

{Figure~\ref{fig:model_vs_fit_amu} compares the individual contributions to \amuL\ from each of the {known} states in $G_{\rm fake}(t)$ (top panel, blue) with those from each state in the two-state fit (bottom panel, red).}
Although the fitted energies are only a compromise between the actual energy levels, the value of \amuL\ obtained from the fitted correlator (even with $t^*=0$~fm) agrees with the known value to $< 2 \times 10^{-10}$. This is because the $t$-dependence of the fit correlator tracks $G_\mathrm{fake}(t)$ closely over the region of~$t$ that matters to~\amuL; the data are not
sufficiently precise to distinguish between a two-state theory and the real theory.
We have repeated this test using model spectra corresponding to each of our lattice spacings $a\approx 0.15$--$0.06$~fm, and find the same conclusions.
This indicates that our simple fit ansatz with two or more exponentials is sufficient to obtain the correct \amuL\ to within the quoted statistics $\oplus$ fit uncertainties.

\begin{figure}[tb]
\centering
\includegraphics[width=0.4\textwidth]{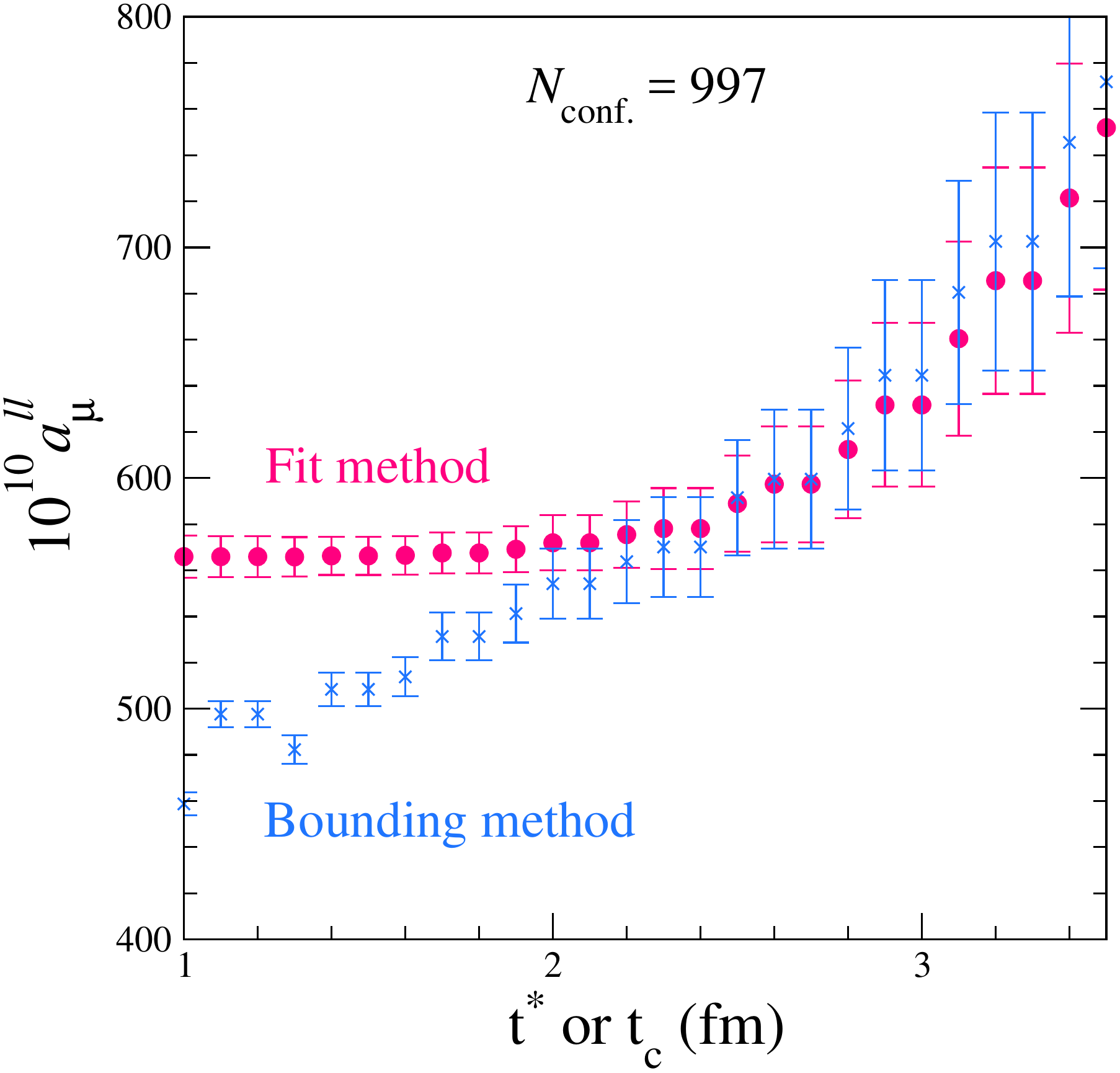}
\caption{Comparison of \amuL\ from our noise-reduction strategy with BMW's bounding method~\cite{Borsanyi:2016lpl} on the low-statistics $a \approx 0.15$~fm ensemble.  The $x$-axis shows either the value of $t^*$ employed in the data+fit method or of $t_c$ used in the bounding method. For the bounding method, we plot the average of \amuL\ obtained from the upper and lower bounds; the two bounds meet at $t_c \sim 2.5$~fm. }
\label{fig:amu_vs_tcut}
\end{figure}

We also compare our approach with the bounding method used by the BMW Collaboration in Ref.~\cite{Borsanyi:2016lpl}. With this approach, they select a value $t_c$ at which they replace the correlator data with the upper bound from a single exponential with the lowest-lying noninteracting two-pion energy level and a lower bound of zero.  They then calculate \amuL\ using the upper and lower bounds on the correlators varying the value of the matching point $t_c$.  They find that the upper and lower bounds meet at around 2.5--3~fm for their data, and take the average of \amuL\ from the upper and lower bounds with $t_c \sim 3$~fm in their recent analysis~\cite{Borsanyi:2017zdw}.  Figure~\ref{fig:amu_vs_tcut} compares \amuL\ computed with our fit method and with BMW's bounding method on the low-statistics $\approx 0.15$~fm ensemble.
(See Table~\ref{tab:ensembles} for the relevant energy levels.)
The results obtained with the two approaches agree, but the fit method yields smaller statistical errors on \amuL .  This is because \amuL\ from the fit method is stable for $t^*$ above 1~fm, whereas the upper and lower bounds do not meet until around 2.5~fm, necessitating a larger value for $t_c$.  The consistency between the two noise-reduction strategies further substantiates our approach of using the fitted correlator at large times, and also indicates that we obtain an accurate result for \amuL\ with $t^*\sim 1$--2~fm.

In Fig.~\ref{fig:amu_vs_tcut}, the value of \amuL\ drifts upward beyond $t^*$ or $t_c$ around 2.5~fm, which corresponds to the time beyond which the correlator data no longer satisfy the model-independent upper bound.  {We observe similar behavior on the other ensembles with only $\sim 1000$ configurations.}  {Thus, both the fit and bounding methods can overestimate \amuL\ with noisy data if the replacement time $t^*$ or $t_c$ is chosen to be too large.}

With the correlator fits in hand, we select the value of $t^*$ where we replace $G_{\rm data}(t)$ with $G_{\rm fit}(t)$ in our calculation of \amuL\,\!.  Plots of \amuL\ versus $t^*$ show that the value of \amuL\ is consistent within errors for $t^*$ between~0.5 and~2.5~fm.  Our choice compromises between minimizing the statistical errors and maximizing the contributions from data.  For simplicity, we select the same value of $t^*=2$~fm for all ensembles, which is larger than the value $t^*=1.5$ used in Ref.~\cite{Chakraborty:2016mwy}.  With our current choice, the data contribution to \amuL\ is greater than 90\% on all ensembles.

\subsection{Lattice corrections and continuum extrapolation}
\label{sec:FVDisc}

Before we extrapolate the values obtained for \amuL\ in Sec.~\ref{sec:amuLat} to zero lattice spacing,
we correct the data for the finite lattice spatial volume and for discretization effects from
the  mass splittings between staggered pions of different tastes. Both effects arise from one-loop
diagrams with $\pi\pi$ intermediate states.  As in Ref.~\cite{Chakraborty:2016mwy}, we calculate
them within an extended chiral perturbation theory that includes pions, $\rho$ mesons, and
photons~\cite{Jegerlehner:2011ti}.  We work to one-pion-loop order, but to all orders in the
leading interactions that couple the $\rho^0$-$\gamma$-$\pi\pi$ channels.
Details of the model calculation can be found in Appendix~B in Ref.~\cite{Chakraborty:2016mwy}.

There are three differences between the numerical calculation of finite-volume corrections in
Ref.~\cite{Chakraborty:2016mwy} and in this work.
The first difference is that the full one-loop finite-volume correction, which included a
piece from quark-disconnected contributions, was applied to the raw \amuL\ in Ref.~\cite{Chakraborty:2016mwy}.
Here we apply the quark-connected part of the one-loop finite-volume correction, which is 10/9 times
the full one-loop value.
Consequently our continuum-limit value of \amuL\ will be larger
than that in Ref.~\cite{Chakraborty:2016mwy}.
We address contributions to \amu\ from quark-disconnected
contributions separately in Sec.~\ref{sec:IBQEDDisc}.

The second difference from Ref.~\cite{Chakraborty:2016mwy} is that here we do not
attempt to correct for differences between the simulated and physical values
of the $\rho$~meson's mass and decay constant by rescaling
contributions to~$a_\mu$. The majority of the lattice ensembles
used in Ref.~\cite{Chakraborty:2016mwy} have pion masses substantially
larger than the physical pion mass; $\rho$~rescaling was used to reduce the dependence
of \amuL\ on the light-quark mass.
 Here, however,
all of our lattice ensembles use light-quark masses that are close
to their physical values.

The third difference from Ref.~\cite{Chakraborty:2016mwy} is a consequence
of the second difference. Without $\rho$ rescaling, we must include
additional finite-volume
corrections coming from the $\rho$'s parameters
(specifically $\Sigma(0)$ in Eqs.~(B20) and~(B22)
in Ref.~\cite{Chakraborty:2016mwy}).
These
new corrections are relatively small,
adding $5\times10^{-10}$ to $13\times10^{-10}$ to~$a_\mu$,
depending upon the lattice ensemble.\footnote{We only include the finite-volume part of this  correction because effects due to the staggered-pion mass splittings have not been calculated (and vanish as $a^2\to0$). Note also that we approximate parameters $\hat m$ and $\hat f$ by the physical $\rho$~mass and decay constant, respectively, in the effective field theory used to calculate this correction (and all other finite-volume corrections); see Appendix~B of~\cite{Chakraborty:2016mwy}.}

For staggered quarks, the sea-pion masses are heavier
than the taste-Goldstone pion for other representations of the
approximate SO(4) taste symmetry.
The taste splittings are discretization errors,
and thus decrease with lattice spacing.
Consequently, the combined finite-volume plus discretization
corrections are largest for our coarsest lattices, and decrease toward the continuum.
The leading finite-volume correction to \amuL\ in chiral perturbation
theory is positive~\cite{Aubin:2015rzx}.
In total, the finite-volume plus discretization corrections to \amuL\
for the lattice ensembles employed in our analysis range from approximately
$68 \times 10^{-10}$ at $a\approx 0.15$~fm to $31 \times 10^{-10}$
at $a\approx 0.06$~fm. These include the leading-order contribution,
from $\pi\pi$~loops, as well as next-to-leading-order
corrections from the pion's charge radius and pion-pion scattering
(see Appendix B of Ref.~\cite{Chakraborty:2016mwy}).
The next-to-leading-order corrections vary from ensemble to ensemble, but are
are smaller than $6\times 10^{-10}$ for our ensembles.
Note that these subleading corrections are not included
in the analyses of BMW~\cite{Borsanyi:2017zdw}
and RBC/UQKCD~\cite{Blum:2018mom}.
They are included, however, in the more recent analysis
of Ref.~\cite{Aubin:2019usy};
our corrections are consistent with theirs (within errors).

\begin{figure}[tb]
\centering
\includegraphics[width=0.4\textwidth]{{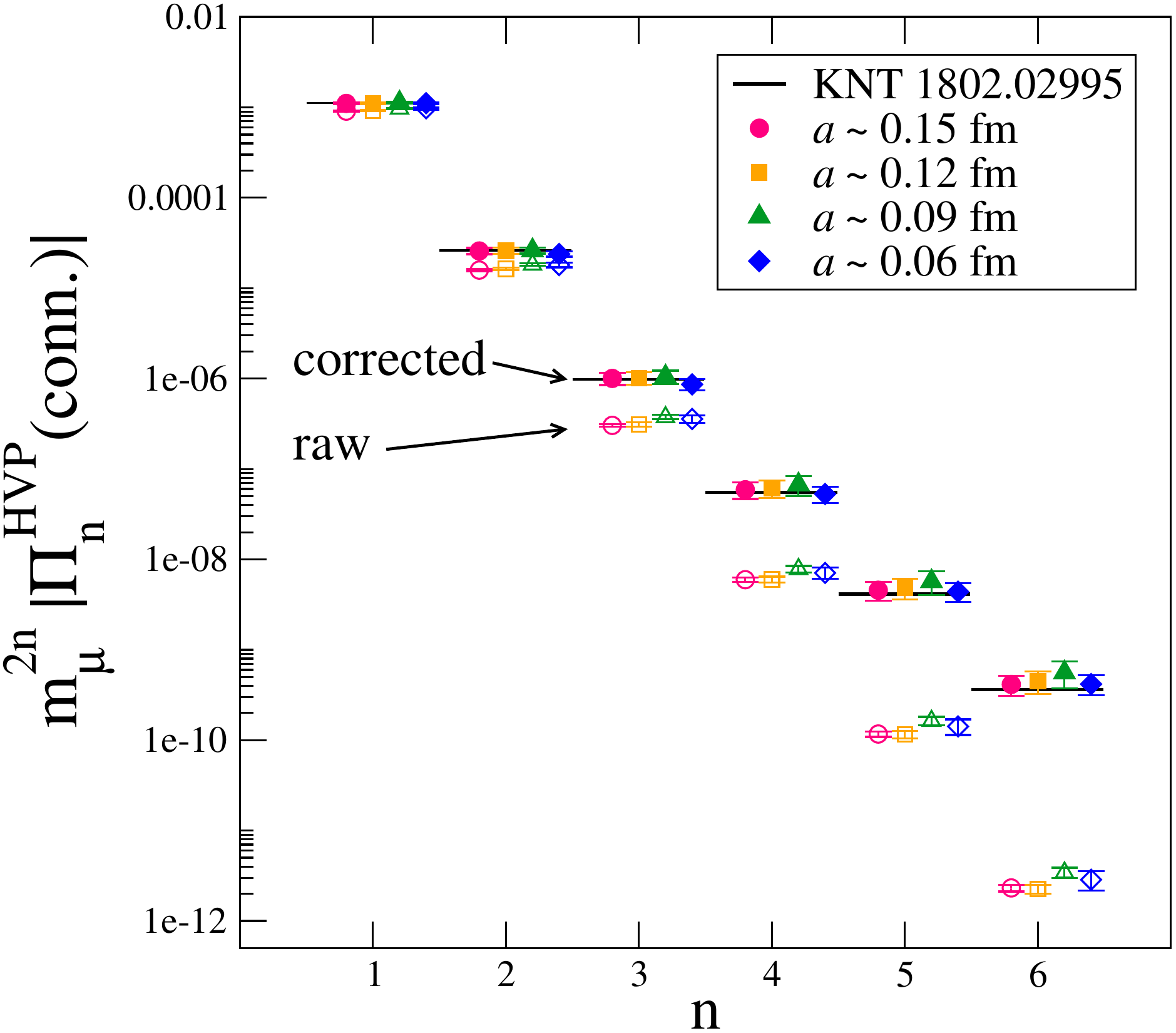}}
\caption{  Quark-connected Taylor coefficients of the renormalized vacuum-polarization function before {(empty)} and after {(filled symbols)} lattice corrections compared with the R-ratio determination from Keshavarzi {\it et al.}~\cite{Keshavarzi:2018mgv}.  We take the $s$-, $c$-, and $b$-quark connected $\Pi_i$s from HPQCD's companion calculations on the MILC HISQ ensembles~\cite{Chakraborty:2014mwa,Colquhoun:2014ica}.}
\label{fig:FVPis}
\end{figure}

As in Ref.~\cite{Chakraborty:2016mwy}, we can test our estimates of the lattice corrections by comparing
our results for the Taylor coefficients of the vacuum-polarization function with phenomenological
determinations from R-ratio data.  Figure~\ref{fig:FVPis} compares our results for the
total quark-connected contributions to $\Pi_1$--$\Pi_6$ before and after the combined finite-volume
plus discretization corrections are applied with a recent phenomenological determination
by Keshavarzi, \emph{et al.}~\cite{Keshavarzi:2018mgv}.  Because the experimental data include
all possible diagrammatic contributions, for this test, we use the full one-loop correction,
which includes both the connected and disconnected pieces. For our full range of lattice
spacings, the corrections bring the lattice-QCD results into agreement with experiment,
up to the 1--2\% level that might be expected from the
small effects of strong-isospin breaking, QED and quark-line
disconnected diagrams missing from our calculation. Note that
the high-$n$ moments demonstrate that the continuum limit of our
chiral theory agrees well with experiment, since the lattice contributions
there are almost negligible (but these moments contribute
little to $a_{\mu}$, as Fig.~\ref{fig:FVPis} also shows).
These comparisons provide strong evidence that our estimated
corrections are reliable both as a function of lattice volume
and as a function of lattice spacing.

In Ref.~\cite{Chakraborty:2016mwy}, this model was also tested by comparison with an explicit finite-volume study on three $a\approx 0.12$~fm ensembles with different spatial volumes but otherwise identical parameters. Because
the pions were unphysically heavy on these lattices, there was little sensitivity to
the spatial volumes.  However, even the small spread in the raw results for \amuL\ of 3(1)\% was removed
by the application of our combined finite-volume plus discretization corrections, providing further confidence in the method.

We can also compare our model with more recent
finite-volume studies based on simulation results. These find finite-volume
shifts of
$\Delta a_\mu^{ll}(\mathrm{conn.})\big(5.4\,\mathrm{fm}\!\to\!10.8\,\mathrm{fm}\big) = 40(18) \times 10^{-10}$,
from the PACS Collaboration~\cite{Shintani:2019gic}, and
$\Delta a_\mu^{ll}(\mathrm{conn.})\big(4.66\,\mathrm{fm}\!\to\!6.22\,\mathrm{fm}\big) = 21.6(6.3) \times 10^{-10}$,
from the RBC/UKQCD Collaboration~\cite{LehnerLat18}. Our model, with
all pion masses $M_\pi=M_{\pi^0}$ and no staggered-pion mass splittings, gives shifts of
$25(4)\times10^{-10}$ and $20(3)\times10^{-10}$, respectively. These estimates
agree with the lattice results above, within their large statistical uncertainties.

Before extrapolating our results for \amuL\ at nonzero lattice spacing to the continuum limit,
we adjust the simulation values for the fact that our pion masses differ by a few MeV between
ensembles (see Table~\ref{tab:ensembles}) and from the physical value.  Using the same chiral model
described above, we remove the continuum quark-connected contribution to \amuL\
from $\gamma \to \pi^+\pi^- \to \gamma$ with the pion mass set equal to the simulation result
for the Goldstone pion (and all other tastes of pion, once lattice artefacts are removed).
We then reintroduce the continuum quark-connected $\pi\pi$ contribution, but with the pion
mass set equal to $M_{\pi^0} =   134.9766(6)$~MeV~\cite{Patrignani:2016xqp}.
Although the shifts are numerically tiny on the ensembles with $M_{\pi_5} \sim 135$~MeV,
the value of \amuL\ on the outlying $a \approx 0.09$~fm ensemble with $M_{\pi_5} \sim 128$~MeV
is decreased significantly, by about $-8 \times 10^{-10}$.

\begin{figure}[tb]
\centering
\includegraphics[width=0.45\textwidth]{{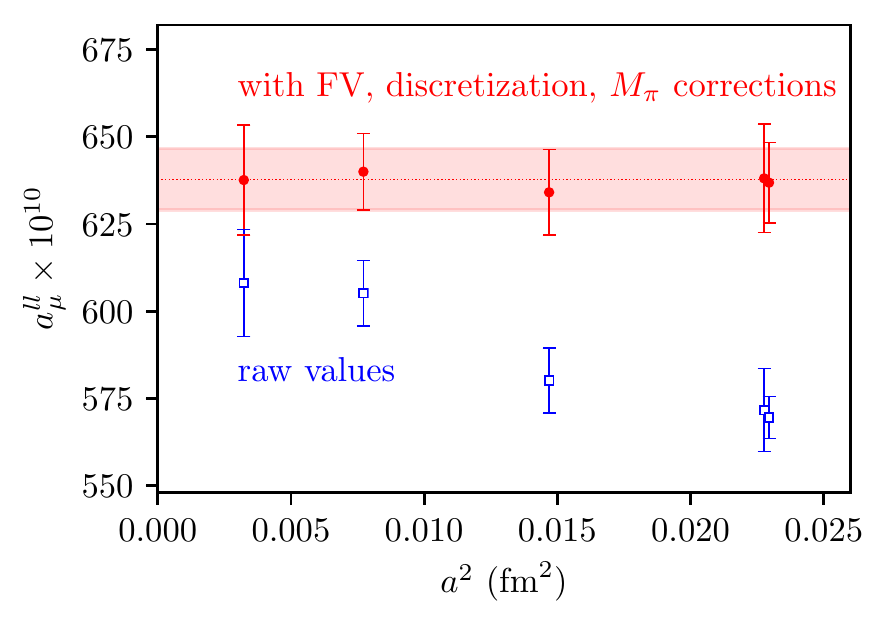}}
\caption{  Lattice-spacing dependence of \amuL\ before (open blue squares) and after (filled red circles) finite-volume, taste-breaking, and $M_\pi$ corrections are applied.  The horizontal light-red band and
red dotted line show the continuum-limit result \amuL~$=\amuLval$ obtained by fitting
the corrected data points with the function in Eq.~(\ref{eq:amu-fitfunction}).}
\label{fig:amu_vs_a2}
\end{figure}

Finally, in order to account for higher-order contributions not included in the corrections,
we assign $15\%$ uncertainties to the net finite-volume and taste-breaking
corrections on both \amuL\  and the Taylor
coefficients. Reference~\cite{Chakraborty:2016mwy} assigned $10\%$ uncertainties to these corrections.
We use a larger uncertainty here because of the new sources of finite-volume error,
associated with $\rho$~parameters, that did not arise in the earlier analysis (see discussion above).
These uncertainties are included in the errors on the corrected results
listed in Table~\ref{tab:amu_corrections} and shown in Fig.~\ref{fig:amu_vs_a2}.

Figure~\ref{fig:amu_vs_a2} shows the lattice-spacing dependence of \amuL\ before and
after both lattice and $M_\pi$ corrections have been applied to the results obtained
in Sec.~\ref{sec:amuLat}, while Table~\ref{tab:amu_corrections} gives the numerical values.
The net corrections range from about
$+11$\% at $a\approx 0.15$~fm to about  $+5$\% at $a\approx 0.06$~fm.
Before corrections, the data display a
large negative slope in $a^2$.
This is quite unlike what was seen for the $s$-quark connected contribution
to \amu\ in Ref.~\cite{Chakraborty:2014mwa}, which
also used the HISQ action and some of the same gauge-field ensembles as we use.
There the variation with lattice spacing, from $a\approx 0.15$~fm to the continuum,
was only 0.5\%. Most lattice-spacing artifacts are larger
for $s$-quarks than for $u/d$-quarks, but taste-splittings are much larger
for pions than for kaons. Hence the large
lattice-spacing dependence seen here, before corrections are made,
are almost certainly due to taste-splittings in the
pion masses~\cite{Chakraborty:2016mwy}.
These should be
greatly reduced by our corrections which account for the
leading effects from taste splitting.
Indeed, Fig.~\ref{fig:amu_vs_a2} shows no
evidence at all of $a^2$~dependence in our corrected data.
The fact that our combined finite-volume and
discretization corrections remove the data's lattice-spacing dependence
is perhaps the strongest evidence that our model for
estimating these effects correctly describes
the physics that underlies our numerical simulations.

\begin{table*}[tb]
    \caption{Light-quark connected contribution to $a_\mu^{\rm HVP}$ and the slope and curvature of the renormalized vacuum polarization before and after applying finite-volume, discretization, and $M_\pi$ corrections.  {Errors shown include uncertainties from statistics, two-point correlator fits, current renormalization, scale-setting, and finite-volume and discretization corrections.  Results on all of the ensembles are correlated through the common scale $w_0$ used to convert $m_\mu$ from physical to lattice units. Results on the two $a \approx 0.15$~fm ensembles are also correlated to a much smaller extent because of the shared renormalization factor, $Z_{V,\bar{s}s}$.}  \vspace{1mm}}
    \label{tab:amu_corrections}
\begin{ruledtabular}
\begin{tabular}{lllllll}
& \multicolumn{2}{c}{$10^{10} a_{\mu}^{ll} ({\rm conn.})$} & \multicolumn{2}{c}{$\Pi_1^{ll}({\rm conn.})$ (GeV$^2$)} & \multicolumn{2}{c}{$\Pi_2^{ll}({\rm conn.})$ (GeV$^4$)} \\
$\approx a$ (fm) & Raw  & Corrected & Raw  & Corrected & Raw  & Corrected \\
\hline
0.15 & 572(12) & 638(15) & 0.0814(18)  & 0.0934(26) & $-$0.1250(54)   & $-$0.216(15)\\
0.15 & 570(6) & 637(11)  & 0.08117(94) & 0.0933(20) & $-$0.1271(30)   & $-$0.217(14)\\\hline
0.12 & 580(9) & 634(12)  & 0.0828(14)  & 0.0928(21) & $-$0.1308(45)   & $-$0.213(14)\\
0.09 & 605(9) & 640(11)  & 0.0868(15)  & 0.0937(18) & $-$0.1463(51)   & $-$0.214(13)\\
0.06 & 608(15) & 638(16) & 0.0871(24)  & 0.0927(25) & $-$0.1438(73)   & $-$0.196(11) \\
\end{tabular}
\end{ruledtabular}
\end{table*}

We extrapolate the corrected values in Fig.~\ref{fig:amu_vs_a2} to the
continuum limit using the following fit function,
which allows for residual $a^2$ and
quark-mass errors beyond the corrections
discussed above:
\begin{align}
a_\mu^{ll}(\mathrm{latt.}) = a_\mu^{ll}(\mathrm{conn.})
\left( 1 + c_s \sum_{f=l,l,s,c} \frac{\delta m_f}{\Lambda}
+ c_{a^2}\frac{(a\Lambda)^2}{\pi^2}
\right) \,.
\label{eq:amu-fitfunction}
\end{align}
Here $\delta m_{f} \equiv m_{f} - m_{f}^\mathrm{phys}$,
and $\Lambda = 0.5$~GeV is of order the QCD scale.
This is similar to the fit function employed in Ref.~\cite{Chakraborty:2016mwy},
except that we no longer include terms to extrapolate in the valence-quark mass
because all of our data are at the physical light-quark mass.
The first term in parentheses adjusts for small sea-quark mass mistuning,
while the second removes residual discretization errors; we employ priors
for the coefficients: $c_s = 0.0(3)$ and $c_{a^2} = 0(1)$.
The values of \amuL\ on each ensemble are statistically independent; we
include in our fit correlations between the two $a\approx 0.15$~fm ensembles
from using the same $Z_V$, and between all ensembles from the common value
of $w_0$ used to convert lattice-spacing units to GeV.

Fitting our full data set to Eq.~(\ref{eq:amu-fitfunction}), we obtain
\begin{align}
a_\mu^{ll}(\mathrm{conn.})
&= \amuLval, \nonumber\\ c_s &=  0.00(30), \nonumber\\  c_{a^2} &=  -0.07(83), \!\!
\label{eq:amu-fit-result}
\end{align}
with a $\chi^2/{\rm d.o.f.} = 0.04$ and $p = 1$.
The fit posteriors for both $c_s$ and~$c_{a^2}$ are consistent with zero,
as expected because of
the corrections applied to the data before extrapolation.
Note that $c_{a^2}=-5(1)$ for the raw values in Fig.~\ref{fig:amu_vs_a2}.

\begin{figure}[tb]
\centering
\includegraphics[width=0.4\textwidth]{{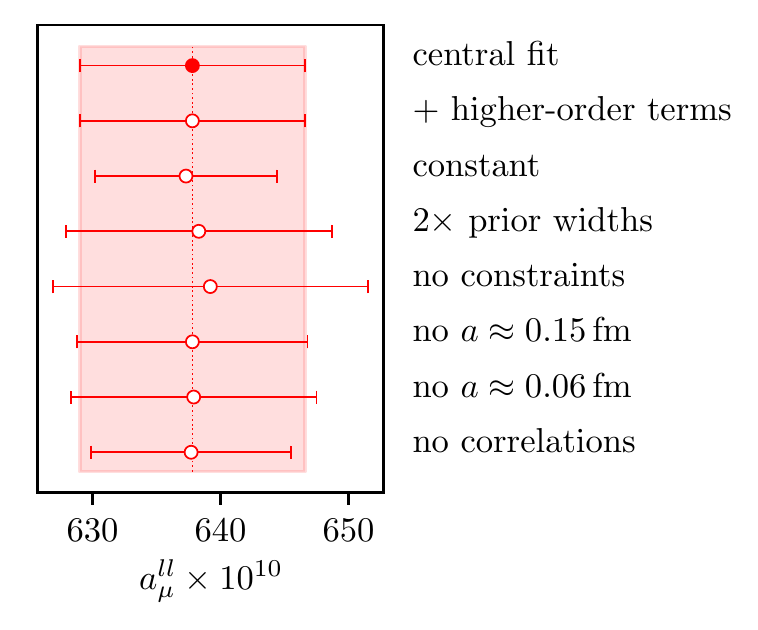}}
\caption{Stability of continuum-limit result for $a_\mu^{ll}(\mathrm{conn.})$
against various fit variations.  From top to bottom, the alternate fits (open circles)
correspond to modifying the central fit (closed circle) by
(i)~adding to the fit function terms proportional to $\chi_f^2$, $\chi_{a^2}^2$,
and $\chi_f\chi_{a^2}$, where $\chi_f \equiv \sum_{f=l,l,s,c} {\delta m_f}/{\Lambda}$
and $\chi_{a^2} \equiv {(a\Lambda)^2}/{\pi^2}$;
(ii)~removing from the fit function the terms proportional to $c_s$ and $c_{a^2}$;
(iii)~doubling the prior widths on all fit parameters;
(iv)~removing constraints on the fit parameters altogether;
(v)~removing data from the coarsest $a$ ($\approx 0.15$~fm);
(vi)~removing data from the finest $a$ ($\approx 0.06$~fm);
and (vii)~removing correlations between the data points.}
\label{fig:amu_fits}
\end{figure}

To study the stability of the values and errors in Eq.~(\ref{eq:amu-fit-result}), we consider a number of fit variations including adding higher-order terms in $a^2$ and $\delta m_{f}$, doubling the prior widths on the fit parameters, and omitting the two coarsest ensembles. We show results for
$a_\mu^{ll}(\mathrm{conn.})$ for several of these variations in Fig.~\ref{fig:amu_fits}.
Most variations differ only slightly from our original fit. The central values
vary by no more than 16\%~of a standard deviation, while
the uncertainties vary by at most 40\%~of a standard deviation.
The fits are excellent, with $\chi^2/\mathrm{d.o.f.}<0.1$ in each case.
The stability exhibited by these results suggest that our fit
error accounts for the systematic uncertainties associated with the
continuum extrapolation. The tiny $\chi^2$~values suggest that
our systematic errors are, if anything, overestimated.

We follow the same procedure to analyze
the slope and curvature of the renormalized vacuum polarization,
first applying finite-volume and taste-breaking discretization corrections,
and then extrapolating to the continuum limit using Eq.~(\ref{eq:amu-fitfunction}).
We obtain continuum-limit values of \PiOneL~=~\PiOneLval\ and \PiTwoL~=~\PiTwoLval.
The $p$ values of the fits are 1.0 and 0.8, respectively.
The fit values for $c_{a^2}$ are $0.02(87)$ and $0.23(98)$ for $\Pi_1$
and $\Pi_2$, respectively; both are consistent with zero and similar to
what we obtained for \amuL\,.
Also the sea-quark mass dependence of $\Pi_1$ and $\Pi_2$ is tiny,
again like~\amuL\,.
Finally, the continuum-limit values \PiOneL\ and \PiTwoL\ are both stable
against the fit variations discussed above for \amuL\,\!.

\section{Results}
\label{sec:Results}

Here we present our final results for \amuL, $\Pi_1^{ll}$, $\Pi_2^{ll}$, and \amu\ and the slope and curvature of $\Pihat(Q^2)$ with comprehensive error budgets.

\subsection{Light-quark connected contribution}
\label{sec:udConn}

Our numerical calculation of \amuL\ and the slope and curvature of the renormalized vacuum-polarization function described in the previous section is with equal up- and down-quark masses, and without electromagnetism.  {These corrections will be included {\it a posteriori}, as is done for other lattice-QCD $g-2$ calculations} in the literature.  It is therefore useful to compare the available lattice-QCD results for \amuL, \PiOneL, and \PiTwoL before putting in the corrections for isospin-breaking and electromagnetism, in order to pin down the source of any disagreements among calculations.

We employ the same definitions for the isospin limit of  \amuL, \PiOneL, and \PiTwoL\ as in Refs.~\cite{DellaMorte:2017dyu,Borsanyi:2017zdw,Blum:2018mom,Giusti:2018mdh}, {which correspond to a world in which all pions have the same mass as the neutral pion.  This allows for a clean comparison among lattice-QCD results.  In Ref.~\cite{Chakraborty:2016mwy}, however, which appeared before Refs.~\cite{DellaMorte:2017dyu,Borsanyi:2017zdw,Blum:2018mom,Giusti:2018mdh}, a different definition was used for \amuL, which we describe below.  Thus, the} result of  Ref.~\cite{Chakraborty:2016mwy} for {this quantity} cannot be directly compared to ours {or} to those of Refs.~\cite{DellaMorte:2017dyu,Borsanyi:2017zdw,Blum:2018mom,Giusti:2018mdh}.

Our results in the isospin-symmetric limit (taken from the fits in the previous section) are
\bea
a_\mu^{ll}(\mathrm{conn.})  & = & \amuLval \times 10^{-10}\, \label{eq:amu_ud} \,, \\
\Pi_1^{ll}(\mathrm{conn.}) &=& \PiOneLmath\, {\rm GeV}^2 \label{eq:Pi1_ud} \,, \\
\Pi_2^{ll}(\mathrm{conn.})  &=&  \PiTwoLmath\,  {\rm GeV}^4 \label{eq:Pi2_ud} \,.
\eea
Table~\ref{tab:udConnEB} gives the breakdowns of the individual error contributions to Eqs.~(\ref{eq:amu_ud})--(\ref{eq:Pi2_ud}).

{We obtain a total uncertainty of 1.4\% on the light-quark connected contribution to \amu\ in the isospin-symmetric limit without electromagnetism.} The largest error contribution to Eq.~(\ref{eq:amu_ud}) comes from the $\sim$~0.5\% uncertainty {on} the scale-setting parameter $w_0$~\cite{Dowdall:2013rya}.  Because the Taylor coefficients of the vacuum-polarization function $\Pi_1$ has dimensions GeV$^{-2}$, the scale-setting error on \amuL\ is approximately twice that of $w_0$.  Statistics, the continuum extrapolation, and finite-volume/discretization corrections {also make significant contributions to the total error.  The remaining contributions
to the uncertainty in \amuL\ are 0.1\% or less.}

\begin{table*}[tb]
    \caption{Error budgets for the $\order(\alpha^2)$ light-quark-connected contribution, the leading Taylor coefficients of the vacuum-polarization function and the muon anomaly in the isospin-symmetric limit without electromagnetism. Sources of uncertainty that were considered, but found to have error contributions $<0.05\%$, are not shown.  \vspace{1mm}}
    \label{tab:udConnEB}
\begin{ruledtabular}
\begin{tabular}{lccc}
Source & \amuL\ (\%) & \PiOneL\ (\%) & \PiTwoL\ (\%) \\[0.5mm]
 \hline
Lattice-spacing ($a^{-1}$) uncertainty & 0.8 & 0.8 & 0.9  \\
Monte Carlo statistics  & 0.7 & 0.8 & 1.2 \\
Continuum ($a\to 0$) extrapolation & 0.7 & 0.7 & 0.8 \\
Finite-volume and discretization corrections  & 0.6 & 0.7 & 2.5 \\
Current renormalization ($Z_V$) & 0.1 & 0.1 & 0.1 \\
Chiral ($m_l$) {interpolation} & 0.1 & 0.1 & {0.0} \\
Sea ($m_s$) adjustment & 0.1 & 0.1 & 0.1 \\
\hline
Total & 1.4\% & 1.{5}\% & {3.1}\% \\
\end{tabular}
\end{ruledtabular}
\end{table*}

In order to compare our result for \amuL\ in Eq.~(\ref{eq:amu_ud}) to the quantity reported in Ref.~\cite{Chakraborty:2016mwy}, we must account for the differences {between} definitions. Instead of {quoting a value at the} neutral pion mass {as we do in this work}, the \amuL\ reported in Ref.~\cite{Chakraborty:2016mwy} {includes the one-loop continuum \pipi\ contribution evaluated at the charged-pion mass.}  In addition, the corrections for finite volume and discretization effects applied in Ref.~\cite{Chakraborty:2016mwy} include the {quark-}disconnected contributions, while the corrections applied here include only the {quark}-connected contributions. The effects of both of these differences increase the value of \amuL\ relative to Ref.~\cite{Chakraborty:2016mwy}.
After accounting for these differences, {however}, our result is {still} about $2\sigma$ higher than the one in Ref.~\cite{Chakraborty:2016mwy}.  This is primarily because we do not rescale the Taylor coefficients {by the ground-state energies of the correlator fits}.

Despite the slightly different meanings of the light-quark connected contribution to \amu\ in Eq.~(\ref{eq:amu_ud}) and in Ref.~\cite{Chakraborty:2016mwy}, it is still useful to compare the error budgets for these quantities.
Compared with that work, we have reduced several key uncertainties.  This is primarily because we employ {\it only} gauge-field configurations with physical-mass light quarks, two of which have finer lattice spacings than in that work. Consequently, the chiral extrapolation, which was an important source of error in Ref.~\cite{Chakraborty:2016mwy}, is replaced here by a chiral interpolation with an associated uncertainty of about 0.1\%. Further, the error due to \Pade\ approximants also made a significant contribution to the total uncertainty in Ref.~\cite{Chakraborty:2016mwy}.  It is reduced here to below 0.05\% by using higher-order [3,2] and [3,3] \Pade\,\!s.  {Two of our uncertainty contributions in Table~\ref{tab:udConnEB}, however, are larger than in Ref.~\cite{Chakraborty:2016mwy}.  Because, in this analysis, we} do not rescale the Taylor coefficients, our quoted lattice-spacing error is about 20~times larger than the estimate in that work.  Our statistical and continuum-extrapolation errors are also two and three times larger, respectively, because the statistical errors increase with decreasing quark mass, and we only employ physical-mass light quarks.   Overall, our total error on \amuL\ is comparable to, but slightly larger than, the 1.1\% error quoted in Ref.~\cite{Chakraborty:2016mwy}.   {Note, however, that we have eliminated two systematic errors present in the result of Ref.~\cite{Chakraborty:2016mwy} that were difficult to estimate, and replaced them with statistical and systematic uncertainties that can be estimated more reliably.

 Figure~\ref{fig:amuudConn} compares our result for \amuL\ in Eq.~(\ref{eq:amu_ud}) with recent unquenched lattice-QCD calculations~\cite{DellaMorte:2017dyu,Borsanyi:2017zdw,Blum:2018mom,Giusti:2018mdh,Shintani:2019wai,Gerardin:2019rua,Aubin:2019usy}.  Our result is compatible with most of the independently obtained values in the literature.  Quantitatively, it agrees well with the published determinations by the BMW and ETM Collaborations~\cite{Borsanyi:2017zdw,Giusti:2018mdh}, with the published results from Mainz (with $N_f$=2)~\cite{DellaMorte:2017dyu} and RBC/UKQCD~\cite{Blum:2018mom}, and with the calculation of Aubin {\it et al.}~\cite{Aubin:2019usy}.
 Our result for \amuL\ is somewhat lower, however, than recent calculations (that appeared after this paper) by Mainz (with $N_f$=3)~\cite{Gerardin:2019rua} and Shintani and Kuramashi~\cite{Shintani:2019wai}.

\begin{figure}[tb]
\centering
\includegraphics[width=0.5\textwidth]{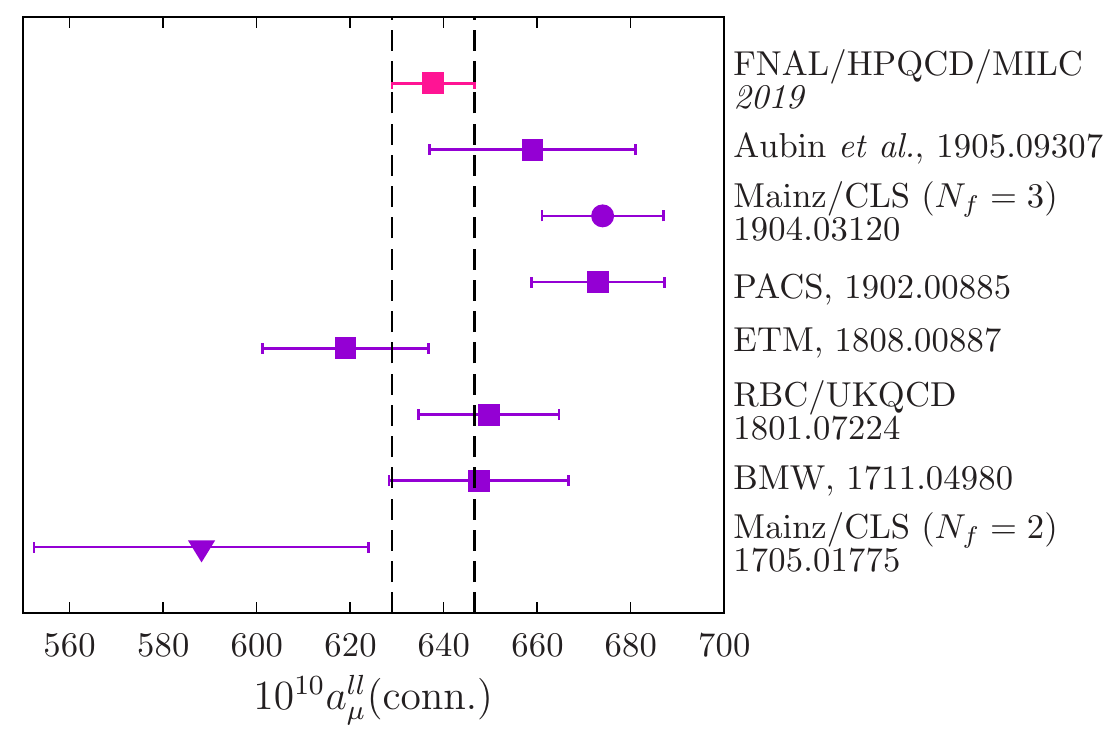}\vspace{-0.1in}
\caption{ Comparison of our result in Eq.~(\ref{eq:amu_ud}) for the light-quark connected contribution to \amu\ with {recent} unquenched lattice-QCD results~\cite{DellaMorte:2017dyu,Borsanyi:2017zdw,Blum:2018mom,Giusti:2018mdh,Shintani:2019wai,Gerardin:2019rua,Aubin:2019usy}.  All values correspond to isospin-symmetric QCD without electromagnetism.  Results for \amuL\ from four-{, three, and two-flavor} QCD simulations are denoted by squares, {circles, and triangles, respectively}.  Note that the RBC/UKQCD Collaboration employed three-flavor QCD gauge-field configurations, and then added the charm sea-quark contribution estimated from perturbation theory {\it a posteriori}.}
\label{fig:amuudConn}
\end{figure}

Finally we discuss the error budgets for the slope and curvature of $\Pihat(Q^2)$, which are also given in Table~\ref{tab:udConnEB}.  The uncertainty breakdown for \PiOneL\ is similar to that for \amuL\ because the two are proportional at lowest order in the Taylor expansion.  The errors for \PiTwoL\ are different because it is
more infrared than the other two quantities---the uncertainty due to uncalculated (higher-order)
finite-volume/discretization contributions dominates all other contributions to the
error budget. We do not quote values for higher-order Taylor coefficients of $\Pihat(Q^2)$ because the estimated errors from finite-volume plus taste-breaking discretization effects are no longer smaller than or commensurate with the contribution from statistics.

Figure~\ref{fig:Pis_ud_compare} compares our results for the slope and curvature of the renormalized vacuum-polarization function in Eqs.~(\ref{eq:Pi1_ud}) and~(\ref{eq:Pi2_ud}) with those from recent lattice-QCD calculations.  Our result for the leading Taylor coefficient, \PiOneL, is consistent with those of the BMW~\cite{Borsanyi:2016lpl}, ETM~\cite{Giusti:2018mdh}, and RBC/UKQCD~\cite{Blum:2018mom} Collaborations. Our result for the second Taylor coefficient, \PiTwoL, agrees with the calculations of ETM and RBC/UKQCD, but is about ${2.0}\sigma$ {larger} in magnitude than that of BMW.  The larger relative spread in \PiTwoL\ values between the collaborations may be due to the variety of approaches used to control the statistical error in the Euclidean vector-current correlator at large times, since higher moments are sensitive to greater times.

\begin{figure}[tb]
\centering
\includegraphics[height=0.275\textheight]{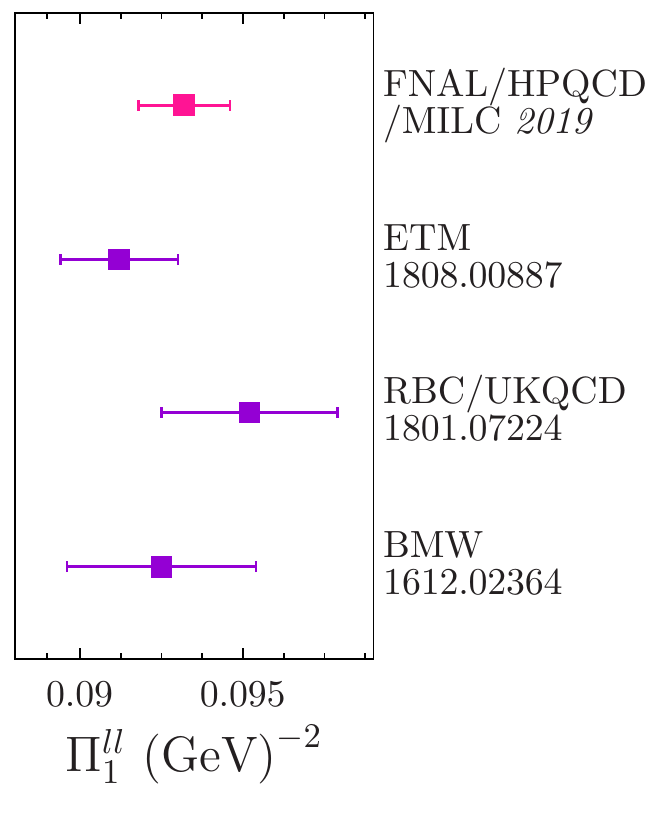} \hspace{-12pt} \includegraphics[height=0.275\textheight]{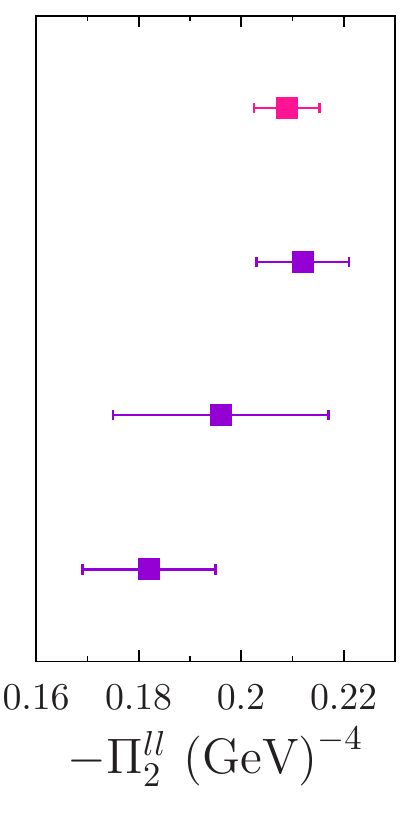} \vspace{-0.15in}
\caption{ Comparison of our results in Eqs.~(\ref{eq:Pi1_ud}) and~(\ref{eq:Pi2_ud}) for the light-quark connected contribution to the slope and curvature of $\Pihat(Q^2)$ with published unquenched lattice-QCD results from Refs.~\cite{Borsanyi:2016lpl,Blum:2018mom,Giusti:2018mdh}.  All values correspond to isospin-symmetric QCD without electromagnetism.  Note that we multiplied the Taylor coefficients quoted in Refs.~\cite{Borsanyi:2016lpl,Giusti:2018mdh} by the charge factor $q_u^2 + q_d^2 = 5/9$ so that they correspond to our normalization convention. }
\label{fig:Pis_ud_compare}
\end{figure}

\subsection{Isospin-breaking, electromagnetic, and quark-disconnected contributions}
\label{sec:IBQEDDisc}

To be able to compare our total summed over all quark flavors with experiment,
we need to correct our result for \amuL\ [Eq.~(\ref{eq:amu_ud})] for contributions due to strong-isospin breaking, QED~effects, and light-quark disconnected contributions.
We will do this in four steps.
First, we will consider these corrections for just diagrams with $\pi\pi$ intermediate~states because they can be calculated reliably from {the} chiral model {used} in Sec.~\ref{sec:FVDisc}.
Next, we will examine separately the remaining corrections from disconnected diagrams, strong isospin breaking, and~QED.
To estimate these contributions, we rely on our own lattice-QCD calculations when
available, models, and phenomenology, and take generous uncertainties to cover roughly the spread of values in the literature.
Table~\ref{tab:QEDSIBDiscCorrections} summarizes our estimates of the corrections to \amu, $\Pi_1^{\rm HVP,LO}$, and $\Pi_2^{\rm HVP,LO}$ from the omission of these effects.

\subsubsection{$\pi\pi$ corrections}
A large part of the isospin, electromagnetic, and quark-disconnected corrections comes from diagrams in Fig.~\ref{fig:HVP} with \pipi\ intermediate states.
These corrections can be estimated using the leading term in our chiral model.
As discussed in Sec.~\ref{sec:FVDisc}, the chiral model gives an excellent description of the finite-volume and taste-breaking discretization effects in our numerical data, and should therefore also be reliable here.

Because of spin-statistics, there is no $\pi^0\pi^0$ contribution to \amu.
Hence the $\pi^0\pi^0$ pieces must cancel between connected and disconnected diagrams.
This leaves purely a $\pi^+\pi^-$ contribution, so it is clear that we should use the $\pi^+$ mass when calculating corrections to our lattice-QCD result for \amu~\cite{Chakraborty:2016mwy}.

In Sec.~\ref{sec:FVDisc}, using our chiral model, we removed the continuum quark-connected contribution to \amuL\ from $\gamma \to \pi^+\pi^- \to \gamma$ with the pion mass set equal to the simulation result for the Goldstone pion, and then reintroduced {it} with the pion mass set equal to $M_{\pi^0}$.
This {is} an artificial choice designed to yield a result for \amuL\ in a world with equal $u$- and $d$-quark masses and without photons.
Now, we can use our chiral model to subtract the continuum quark-connected \pipi\ contribution with the pion mass set equal to $M_{\pi^0}$, and add the quark-connected contribution with the pion mass set equal to $M_{\pi^+} =  139.57018(35)$~MeV~\cite{Patrignani:2016xqp}.
This yields {for the part of} isospin-breaking/electromagnetic correction coming from the $\pi^0\pi^+$ mass
difference
\be
\Delta a_\mu^{\pi\pi}(M_{\pi^0}\to M_{\pi^+})\ = -4.3 \times 10^{-10}\,. \label{eq:pipi_mpi_correction}
\ee
This correction already takes care of some QED effects because the difference between the $\pi^+$ and $\pi^0$ masses comes largely from QED.

Next, we calculate the contribution to \amu\ from quark-disconnected diagrams in Fig.~\ref{fig:HVP} with \pipi\ intermediate states.
Because the $\pi\pi$ contribution appears only in the isospin-1 channel, the ratio of quark-disconnected to quark-connected contributions is $-1/10$ from the ratio of
appropriate quark electric charges~\cite{DellaMorte:2010aq, Chakraborty:2015ugp}.
Therefore the ratio of quark-disconnected to total contributions is $-1/9$.
A calculation of the {\it full} \pipi\ contribution to \amu\ within our chiral model
using the experimental $M_{\pi^+}$ gives $a_{\mu}^{ud}(\pi\pi) = 71 \times 10^{-10}$~\cite{Chakraborty:2016mwy}.
Multiplying this by $-1/9$, we arrive at a quark-disconnected correction from \pipi\ states of
\be
\Delta a_{\mu}^{\pi\pi}({\rm disc.})\ = -7.9 \times 10^{-10} \label{eq:pipi_disc_correction}\,.
\ee

Adding Eqs.~(\ref{eq:pipi_mpi_correction}) and~(\ref{eq:pipi_disc_correction}), we arrive at a total \pipi\ correction to \amu\ from strong-isospin breaking, electromagnetism, and quark-disconnected diagrams of
\be
\Delta a_{\mu}^{\pi\pi}
= -12(3) \times 10^{-10}\,. \label{eq:pipi_total_correction}
\ee
We assign a 25\% {error} to this value because the dominant corrections to the leading-order $\pi\pi$~contribution in our chiral model (from the pion charge radius) enter at this level~\cite{Chakraborty:2016mwy}.
We follow the same prescription to estimate with our chiral model the \pipi\ corrections to the slope and curvature of $\Pihat(Q^2)$.

\subsubsection{Residual light-quark disconnected corrections}

There are also quark-line disconnected corrections to \amu\ that have nothing to do with the $\pi\pi$~contribution discussed above.
Following the approach of Chakraborty~\textit{et al.}, we estimate these by examining the contributions to the anomaly from the~$\rho$ and $\omega$~mesons~\cite{Chakraborty:2015ugp}.  Together, these two resonances account for almost 80\% of the total~\amu~\cite{Jegerlehner:2017lbd, Davier:2017zfy, Keshavarzi:2018mgv}.

The ratio of the disconnected to connected moments coming from the~$\rho$ and~$\omega$ is given by Eq.~(11) in Ref.~\cite{Chakraborty:2015ugp}:
\be
\left.\frac{(\Pi_{j})_D}{(\Pi_{j})_C}\right|_{\mathrm{res}}
\approx \frac{1}{10}\left[\frac{m_{\rho}^{2j+2}f_{\omega}^2}{m_{\omega}^{2j+2}f_{\rho}^2}-1 \right],
\ee
where the moments (now) include the quarks' electric charge factors.
This relation, when combined with experimental data for $\rho$ and $\omega$ masses and bounds on their widths, implies a disconnected contribution from non-$\pi\pi$ states of
\be
\Delta a_{\mu}^\mathrm{\rho\omega}({\rm disc.}) = -5(5) \times 10^{-10}\,, \label{eq:rho_disc_correction}
\ee
where the error is from the uncertainty on the inputs.  The correction in Eq.~(\ref{eq:rho_disc_correction}) does not include disconnected diagrams that mix light-quark and $s$-quark loops (connected to the photons), but these are known to be much smaller~\cite{Chakraborty:2015ugp}.  Again, we estimate the disconnected contribution from the $\rho$ and $\omega$ resonances to the Taylor coefficients $\Pi_1$ and $\Pi_2$ in the same manner.

Note that adding the above $-5(5)\times 10^{-10}$ to the $\pi\pi$ contribution from Eq.~(\ref{eq:pipi_disc_correction}) gives $-13(5) \times 10^{-10}$ for the total quark-line
disconnected contribution.
This is well in line with direct lattice-QCD calculations of the quark-disconnected contribution to \amu\ in the isospin-symmetric limit and without QED---including $s$-quark contributions, the BMW Collaboration finds $a_{\mu}^{\rm HVP,LO}(\rm disc.) = -12.8(1.9) \times 10^{-10}$~\cite{Borsanyi:2017zdw}, while RBC/UKQCD obtains $a_{\mu}^{\rm HVP,LO}(\rm disc.) = -11.2(4.0)\times 10^{-10}$~\cite{Blum:2018mom}---and further supports the reliability of our model calculations.

\subsubsection{Residual strong-isospin breaking corrections}

The effects from QCD-isospin breaking ({\it i.e.}, quark-mass differences) and QED are intertwined both in nature and in lattice-QCD simulations because QED~contributions shift the bare quark masses.
Here we \textit{define} the residual strong-isospin correction to \amu\ as the shift relative to the isospin-symmetric value \amuL\ that results when the bare~$u$ and~$d$ quark masses are retuned separately so that
(i)~their average gives the experimental value for the $\pi^0$ mass [as required for \amuL],
and (ii)~their ratio has the physical value obtained from lattice-QCD calculations including electromagnetism~\cite{Basak:2016jnn,Basak:2018yzz}.
Note that $\pi\pi$~contributions largely cancel in this correction because the pion mass is primarily sensitive to the average light-quark mass.

There has been much recent work using lattice-QCD simulations to estimate the strong-isospin breaking correction to \amu.
Our first calculation of these corrections considered quark-line connected diagrams only on a relatively coarse lattice spacing, but employed physical light-quark masses~\cite{Chakraborty:2017tqp}.
We found a relative correction of $\delta a_{\mu}^{{\rm HVP,LO}}({\rm SIB}) =$+1.5(7)\%, which translates into an absolute correction $\Delta a_\mu^{{\rm HVP,LO}}({\rm SIB}) = +9.5(4.5)\times 10^{-10}$ when combined with \amuL\ from Eq.~(\ref{eq:amu_ud}).
Subsequent results from the RBC/UKQCD Collaboration of $+10.6(8.0)\times 10^{-10}$~\cite{Blum:2018mom},
and by the ETM Collaboration of {$+6.0(2.3)\times 10^{-10}$~\cite{Giusti:2019xct}} (taking the continuum limit from three lattice-spacing values), are in good agreement.

When only the quark-line connected diagrams are considered, the strong-isospin breaking correction will contain unphysical effects from $\pi\pi$ states where the $\pi$ meson is composed of $u\overline{u}$ and $d\overline{d}$ states.
These effects will be positive since isospin-breaking effects are positive and the
 ``'$\pi_u$'' meson is unnaturally light. They will be canceled, as discussed above, when the quark-line disconnected diagram is included.
This means that we might expect substantial negative contributions from the quark-line disconnected diagrams, relative to the isospin-symmetric case, when strong-isospin breaking effects are included.
Indeed, our preliminary results for the strong-isospin-breaking correction to the quark-disconnected contribution confirm this~\cite{CDSchwinger18}.
We therefore increase the errors on our initial estimate of the total residual correction from strong-isospin breaking (from~\cite{Chakraborty:2017tqp}) to allow for disconnected contributions {of a commensurate size}, giving
\begin{equation}
   \Delta  a_\mu^{ud}(\mathrm{SIB})
   = 10(10)\times 10^{-10}\,. \label{eq:resSIB}
\end{equation}

The analysis in Ref.~\cite{Chakraborty:2017tqp} also yielded estimates for the strong-isospin breaking corrections to the Taylor coefficients of $\Pihat(Q^2)$ of $\delta \Pi_1^{{\rm HVP,LO}}(m_u\ne m_d) =$+1.6(6)\% and $\delta \Pi_2^{{\rm HVP,LO}}(m_u\ne m_d) =$+3.0(8)\%.  We employ these values to obtain the absolute corrections to $\Pi_1^{\rm HVP,LO}$ and $\Pi_2^{\rm HVP,LO}$, and again increase the uncertainties to 100\% to allow for large quark-disconnected contributions.

\subsubsection{Residual QED corrections}

We have already included a sizeable part of the full QED correction by replacing
the $\pi^0$~mass by the $\pi^+$~mass in the $\pi\pi$~contribution.
We estimate the residual corrections from QED, beyond those accounted for above,
via power-counting to be of order $\alphaEM \sim 1\%$.  This yields an estimate for the absolute correction to \amu\ of
\begin{equation}
    \Delta a_\mu^{ud}(\mathrm{QED})
    = 0(5) \times 10^{-10}\,, \label{eq:resQED}
\end{equation}
where we have taken a central value of zero because we do not know the sign of the correction. We take the same {\it relative} QED error for the Taylor coefficients $\Pi_1$ and $\Pi_2$.

Our estimate of residual QED corrections is consistent with results from the analysis of \amu\ based upon experimental data on $e^+e^- \rightarrow {\rm hadrons}$.
For example, the contribution from the simplest photon channel, $e^+e^- \rightarrow \pi^0\gamma$, is $4.5\times 10^{-10}$~\cite{Keshavarzi:2018mgv}.
Equation~(\ref{eq:resQED}) is also consistent with (still early) efforts to estimate the QED contribution using lattice-QCD simulations~\cite{Blum:2018mom,Giusti:2018vrc,Giusti:2019xct}.
The RBC/UKQCD Collaboration finds $\Delta a_{\mu}^{\rm HVP,LO}({\rm QED}) \approx -1(6)\times 10^{-10}$ from summing results from connected and disconnected diagrams~\cite{Blum:2018mom}, while the ETM Collaboration finds $\Delta a_{\mu}^{\rm HVP,LO}({\rm QED,conn.}) = {1.3(1.0)} \times 10^{-10}$ from connected diagrams only~\cite{Giusti:2019xct}.

\subsubsection{Total contribution from $u/d$ quarks}
Summing the corrections from Eqs.~(\ref{eq:pipi_total_correction}) and
(\ref{eq:rho_disc_correction}--\ref{eq:resQED}) we obtain for the total correction from strong-isospin breaking, QED, and quark-disconnected contributions:
\begin{align}
    \Delta a_\mu^{ud}({\rm SIB,QED,disc.)} &= -7(13)\times 10^{-10}. \label{eq:Delta_amu_ud}
\end{align}
Adding this to \amuL\ [Eq.~(\ref{eq:amu_ud})], we obtain the total contribution to \amu\  from light quarks:
\begin{align}
    a_{\mu}^{ud} & = 630.8(8.8)(13) \times 10^{-10}\,, \label{eq:HVPLOud}
\end{align}
where the first error is from~\amuL\ and the second is from~$\Delta a_\mu^{ud}$.

\begin{table*}[tb]
    \caption{
Summary of our estimates of the corrections to \amu, $\Pi_1^{\rm HVP,LO}$, and $\Pi_2^{\rm HVP,LO}$ from the omission of strong-isospin breaking, QED and light-quark disconnected diagrams.
\vspace{1mm}}
    \label{tab:QEDSIBDiscCorrections}
\begin{ruledtabular}
\begin{tabular}{lY{3.4}d{2.8}d{2.6}}
Contribution & \multicolumn{1}{c}{$10^{10} a_\mu^{ll}({\rm conn.})$} & \multicolumn{1}{c}{$\Pi_1^{ll}({\rm conn.})\ ({\rm GeV}^{-2})$} & \multicolumn{1}{c}{$\Pi_2^{ll}({\rm conn.})\ ({\rm GeV}^{-4})$} \\
\hline
$M_{\pi^0} \to M_{\pi^+}$ & -4..3 & -0.00075 & 0.0057 \\
$\pi\pi$ disconnected & -7..9 & -0.00120 & 0.0044 \\
\hline
Total $\pi\pi$ & -12.(3) & -0.0020(5) & 0.010(3) \\
\hline
$\rho,\omega$ disconnected & -5.(5) & -0.0008(8) & 0.002(1) \\
Strong-isospin breaking & 10.(10) & 0.0015(15) & -0.006(6) \\
Electromagnetism & 0.(5) & 0.0000(6) & 0.000(2) \\
\hline
Total correction & -7.(13) & -0.0013(19) & 0.006(7) \\
\end{tabular}
\end{ruledtabular}
\end{table*}

\subsection{Total leading-order contribution}
\label{sec:LOTot}

Finally, to obtain the total leading-order hadronic vacuum polarization contribution to $a_\mu$, we add the contributions from heavy flavors to $a_\mu^{ud}$ (Eq.~\ref{eq:HVPLOud}).
We take the connected results for strange, charm, and bottom quarks calculated by the HPQCD Collaboration~Ref.~\cite{Donald:2012ga,Chakraborty:2014mwa,Colquhoun:2014ica}.\footnote{{The present author list overlaps with those of Refs.~\cite{Donald:2012ga,Chakraborty:2014mwa,Colquhoun:2014ica}.}}
Disconnected contributions from these quarks are expected to be negligible
compared with our other uncertainties. We follow the same procedure for the Taylor coefficients of the renormalized vacuum-polarization function.

Table~\ref{tab:amu-flavor} gives the individual flavor contributions to \amu, $\Pi_1^{\rm HVP,LO}$, and $\Pi_2^{\rm HVP,LO}$.
More than 90\% of the central value comes from the light-quark connected contribution, as does about 30\% of the error.
The remainder of the error on \amu\ comes from the uncertainty on our estimate of the missing contributions from QED, strong-isospin breaking, and quark-disconnected diagrams.
The contributions from $s, c$, and $b$ quarks generate the remaining $\sim 10\%$ of the central value,
while contributing a negligible amount, $\sim$0.1\%, to the error.
\begin{table*}[tb]
    \caption{Individual flavor contributions to the leading Taylor coefficients of the vacuum-polarization function and the muon anomaly.
    The first error quoted for the $u/d$ contributions is from the lattice analysis; the second comes from
    uncertainties in our estimates of the effects of strong isospin-breaking, electromagnetism, and
    quark disconnected diagrams. Results for strange and heavier quarks include only
    the quark-connected contributions and are not new, but come from earlier HPQCD calculations~\cite{Donald:2012ga,Chakraborty:2014mwa,Colquhoun:2014ica};
    disconnected contributions are expected to be negligible.
    The definitions of the Taylor coefficients include the factor of the quark's electric charge squared.  \vspace{1mm}}
    \label{tab:amu-flavor}
\begin{ruledtabular}
\begin{tabular}{lY{4.10}d{2.12}d{2-12}}
Contribution & \multicolumn{1}{c}{$10^{10} \amumath$} & \multicolumn{1}{c}{$\Pi_1^{\rm HVP,LO} ({\rm GeV}^{-2})$} & \multicolumn{1}{c}{$\Pi_2^{\rm HVP,LO} ({\rm GeV}^{-4})$} \\[1pt]
\hline
Light & 630..8(8.8)(13) & 0.0919(14)(19) & -0.2029(64)(71) \\
Strange & 53..40(60) & 0.007291(78) & -0.00587(12) \\
Charm & 14..40(40) & 0.001840(49) & -0.0001240(43) \\
Bottom & 0..270(40) & 0.0000342(48) & -2.28(37)e-07 \\[1pt]
\hline
Total & 699.(15) & 0.1011(24)  & -0.2089(95)
\end{tabular}
\end{ruledtabular}
\end{table*}

\section{Summary and outlook}
\label{sec:conclusions}

Our results for the leading-order HVP contributions to $a_\mu$ and the slope
and curvature of the renormalized vacuum polarization function are
(Table~\ref{tab:amu-flavor})
\bea
10^{10} \amumath & = &  \amuTOTval \label{eq:amuHVPLO} \\
\Pi_1^{\rm HVP,LO} & = &  0.1011(24)_{u,d}(1)_{s,c,b} {\rm\ GeV}^{-2} \label{eq:Pi1HVPLO} \\
\Pi_2^{\rm HVP,LO} & = &   -0.2089(95)_{u,d}(1)_{s,c,b} {\rm\ GeV}^{-4} \label{eq:Pi2HVPLO}\,
\eea
The total uncertainty on \amu\ is $\sim 2.2\%$, and is dominated by our conservative estimate of the combined
uncertainty from the omission of strong isospin-breaking, electromagnetism, and quark-disconnected contributions in our calculation of the $u/d$-quark contribution (see Sec.~\ref{sec:IBQEDDisc}.).

We also reiterate the key intermediate result of this work, which is our new determination
of the light-quark connected contribution to \amu\ in the isospin-symmetric limit
and without electromagnetism (from Eq.~(\ref{eq:HVPLOud})):
\be
10^{10}  a_\mu^{ll}(\mathrm{conn.})  = \amuLval \,. \label{eq:amuud_repeat}
\ee
This result improves upon, and supersedes the calculation of Chakraborty {\it et al.} in Ref.~\cite{Chakraborty:2016mwy}.   As can be seen from Fig.~\ref{fig:amuudConn}, our determination of \amuL\ has smaller errors than other recent unquenched lattice-QCD calculations~\cite{DellaMorte:2017dyu,Borsanyi:2017zdw,Blum:2018mom,Shintani:2019wai,Giusti:2018mdh,Gerardin:2019rua,Aubin:2019usy}.  This is primarily because our fit method for controlling the statistical errors in the Euclidean vector-current correlator at large times yields smaller uncertainties on \amuL\ than approaches used by other collaborations.

Figure~\ref{fig:amuSummary} compares our determination of the total, leading-order hadronic-vacuum-polarization contribution to $a_\mu$ in Eq.~(\ref{eq:amuHVPLO}) with other lattice-QCD calculations~\cite{Burger:2013jya,DellaMorte:2017dyu,Borsanyi:2017zdw,Blum:2018mom,Giusti:2018mdh,Shintani:2019wai,Gerardin:2019rua} and  phenomenological analyses of experimental $R$-ratio data~\cite{Benayoun:2015gxa,Jegerlehner:2017lbd,Davier:2017zfy,Keshavarzi:2018mgv}.   Our result agrees with all {but one} of the independent lattice calculations, and has a comparable error.\footnote{{As we were finishing this paper, Shintani and Kuramashi presented a new determination of $10^{10} \amumath = 737(+16,-21)$~\cite{Shintani:2019wai} that is $1.5\sigma$ above our result, and is in more than $2\sigma$-tension with the $R$-ratio analyses.}}  It also agrees with the $R$-ratio analyses, although with roughly five to seven times larger uncertainties.

\begin{figure}[tb]
\centering
\includegraphics[width=0.48\textwidth]{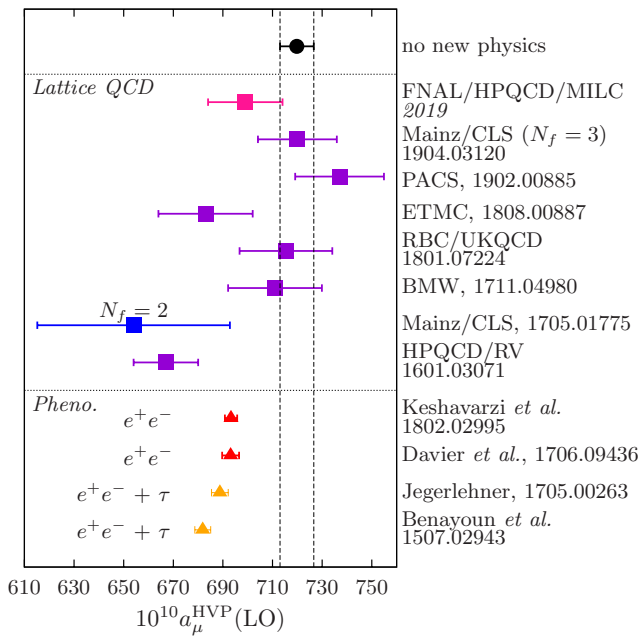}
\caption{  Comparison of our result in Eq.~(\ref{eq:amuHVPLO}) for the leading-order hadronic-vacuum-polarization contribution to the muon anomalous magnetic moment (magenta square) with results from $N_f \geq 2$ lattice QCD~\cite{Burger:2013jya,DellaMorte:2017dyu,Borsanyi:2017zdw,Blum:2018mom,Giusti:2018mdh,Shintani:2019wai,Gerardin:2019rua} (blue and purple squares), and from experimental $e^+e^-$ cross-section data~\cite{Benayoun:2015gxa,Jegerlehner:2017lbd,Davier:2017zfy,Keshavarzi:2018mgv} (red and orange triangles).  The filled black circle shows the value of $a_\mu^{\rm HVP, LO}$ that is implied by the measurement of $a_\mu$ by BNL experiment E821~\cite{Bennett:2006fi} assuming no contributions beyond the Standard Model; vertical dashed lines denote the $\pm 1\sigma$ range~\cite{Chakraborty:2016mwy}.}
\label{fig:amuSummary}
\end{figure}

We also compare our result for \amu\ in Eq.~(\ref{eq:amuHVPLO}) to the expectation from experiment.
Assuming that there are no contributions to the muon anomalous magnetic moment from physics
beyond the Standard Model, the BNL E821 Experiment~\cite{Bennett:2006fi} {implies a value for} \amu\ {of} $720(7)\times 10^{-10}$.  This value is obtained by subtracting from experiment the {calculated} values of QED~\cite{Aoyama:2012wk}, electroweak~\cite{Gnendiger:2013pva} and higher order HVP~\cite{Hagiwara:2011af, Kurz:2014wya} {contributions and the consensus value for the} hadronic light-by-light {term}~\cite{Prades:2009tw}.  Our result is $\amuDIFF\sigma$ below the ``no new physics" value, with about twice the uncertainty.

Clearly the theoretical error on \amu\ in Eq.~(\ref{eq:amuHVPLO}) is still too large to draw any conclusions regarding the presence of new physics, and must be reduced by around a factor of 10 to reach the 0.2\% target precision of the Muon $g-2$ Experiment.  Three key ingredients are still missing from our calculation of \amu\ described here: the effect of the difference between the $u$- and $d$-quark masses and of the quarks' electric charges on the light-quark connected contribution, and the contribution to the total from quark-disconnected diagrams involving $u$, $d$, $s$, and $c$ quarks.  Work on all of these is in progress~\cite{Chakraborty:2017tqp,Yamamoto:2018cqm}. Because they are all small corrections, however, relatively high accuracy is not needed. Ultimately calculations will be done on gluon-field configurations in which the sea quarks have both color and electric charges. Generation of such an ensemble is underway~\cite{YuzhiLat18}.

We must also further reduce the uncertainty on the light-quark connected contribution \amuL\ in Eq.~(\ref{eq:amuud_repeat}). The error budget (Table~\ref{tab:udConnEB}) is dominated by the lattice-spacing uncertainty, statistical errors and the continuum extrapolation. The last two can be reduced by increasing statistics, so that the results at each lattice spacing value are more precise, and hence provide better constraints on the continuum extrapolation.  We have demonstrated here that a calculation with nearly 0.5 million correlators (our
high-statistics sample at $a$ = 0.15fm) resolves issues around how to handle statistical uncertainties at large Euclidean times. Such a sample is numerically expensive to obtain on finer lattices, although tripling the statistics is certainly feasible using the truncated solver method. We estimate that this would reduce our total uncertainty to 1\%.
{Further improvements may be achieved by analyzing additional correlation functions that include two-pion operators to better resolve the large-time behavior of the vector-current correlation functions~\cite{Gerardin:2018sin,MeyerLat18}.}
To get below 1\% requires a reduction in the uncertainty on the {physical value of $w_0$ that determines the} lattice
spacing ({$w_0/a$ is determined very precisely, see Table~\ref{tab:ensembles}}).
This uncertainty {currently} relies on a determination of the pion decay constant, $f_{\pi}$, on the lattice~\cite{Dowdall:2013rya}. The error budget in~\cite{Dowdall:2013rya} shows that the dominant uncertainties are related to
statistical precision and extrapolation to the physical point where $w_0f_{\pi}$ is fixed against
experiment ({assuming a value of $V_{ud}$ from nuclear physics}).
An improvement by a factor of~2 in this uncertainty seems feasible with the higher statistics
gluon-field ensembles now available with physical $m_{u/d}$ on finer lattices.
Analysis on QCD+QED gluon field ensembles will be important here too
to take into account fully the fact that the decaying pion is an electrically charged particle. We also plan to investigate other quantities for determining the lattice spacing.

Given the above discussion, a reduction in uncertainty on the lattice-QCD result for the hadronic vacuum polarization contribution to the muon $g-2$ to $\approx$ 0.5\% is certainly feasible on the timescale of the new experiments. This would give precision comparable to that currently available from using experimental information on $e^+e^- \rightarrow {\rm hadrons}$ and
would allow lattice-QCD results to play a significant role in the unfolding story of the search for new physics in the anomalous magnetic moment of the muon.

\begin{acknowledgments}

We thank Bob Sugar for his scientific leadership and tireless efforts to obtain computational resources, without which the MILC physics program would never have been realized. We thank Bipasha Chakraborty and Jonna Koponen for generating data employed in this analysis.  We thank Alex Keshavarzi, Christoph Lehner, Kotaroh Miura, and Aaron Meyer for useful discussions, and the latter three for providing additional information on the BMW and RBC/UKQCD Collaborations' calculations. We thank M.~Golterman, K.~Maltman, and S.~Peris for useful discussions regarding the high-order finite volume effects calculated in Ref.~\cite{Aubin:2019usy}.

Computations for this work were carried out with resources provided by the USQCD Collaboration, the National
Energy Research Scientific Computing Center and the Argonne Leadership Computing Facility, which are funded
by the Office of Science of the U.S.\ Department of Energy;
and on the DiRAC Data Analytic System at the University of Cambridge, operated by the University of Cambridge High Performance Computing Service on behalf of the U.K. STFC DiRAC HPC Facility, funded by the Department of Business, Innovation and Skills national e-infrastructure and STFC capital grants and STFC Dirac operations grants.
{This work used the Extreme Science and Engineering Discovery Environment (XSEDE) supercomputer Stampede 2 at the Texas Advanced Computing Center (TACC) through allocation TG-MCA93S002.  The XSEDE program~\cite{xsede} is supported by the National Science Foundation under Grant No.\ ACI-1548562.}
{Computations on the Big Red II+ supercomputer were supported in part by Lilly Endowment, Inc., through its support for the Indiana University Pervasive Technology Institute.} the parallel file system employed by Big Red II+ is supported by the National Science Foundation under Grant No.\ CNS-0521433.
This work utilized the RMACC Summit supercomputer, which is supported by the National Science Foundation (Grants No.\ ACI-1532235 and No.\ ACI-1532236), the University of Colorado Boulder, and Colorado State University. The Summit supercomputer is a joint effort of the University of Colorado Boulder and Colorado State University.
This research is part of the Blue Waters sustained-petascale computing project, which is supported by the National Science Foundation (Grants No.\ OCI-0725070 and No.\ ACI-1238993) and the state of Illinois. Blue Waters is a joint effort of the University of Illinois at Urbana-Champaign and its National Center for Supercomputing Applications.

This work was supported in part by the U.S.\ Department of Energy under Awards
{No.~DE-AC02-07CH11359~(T.P.)},
No.~DE-FG02-13ER41976~(D.T.),
No.~DE{-}SC0009998~(J.L.),
No.~DE{-}SC0010005~(E.T.N.),
No.~DE{-}SC0010120~(S.G.),
and No.~DE{-}SC0015655~(A.X.K.);
by the U.S.\ National Science Foundation under Grants
No.\ PHY12-12389~(Y.L.),
No.\ PHY13-16222~(G.P.L.),
No.\ PHY{17-19626}~(C.D., A.V.),
and No.\ PHY14-17805~(J.L.);
by the U.K. STFC under Grants
No.\ ST/L000466/1 and No.\ ST/P000746/1 (C.T.H.D., D.H.),
and No.\ ST/N005872/1 and No.\ ST/P00055X/1 (C.M.);
by the MINECO (Spain) under Grant No.\ FPA2016-78220-C3-3-P~(E.G.);
by the Junta de Andaluc{\'i}a (Spain) under Grant No.~FQM-101~(E.G.);
by the Fermilab Distinguished Scholars Program~(A.X.K.);
by the German Excellence Initiative and the European Union Seventh Framework Program under Grant No.~291763 as well as the European Union's Marie Curie COFUND program~(A.S.K.);
and by the Blue Waters PAID program~(Y.L.).

This document was prepared by the Fermilab Lattice, HPQCD, and MILC Collaborations using the resources of the Fermi National Accelerator Laboratory (Fermilab), a U.S. Department of Energy, Office of Science, HEP User Facility.
Fermilab is managed by Fermi Research Alliance, LLC (FRA), acting under Contract No.\ DE-AC02-07CH11359.

\end{acknowledgments}

\bibliographystyle{apsrev4-1} 
\bibliography{./bibliography}

\end{document}